\documentclass[11pt,a4paper]{article}

\pdfoutput=1
\usepackage{jheppub}

\usepackage{color}
\usepackage{appendix}
\usepackage{graphicx}
\usepackage{wrapfig,enumerate,slashed}
\usepackage{footmisc}
\usepackage{amsmath}
\usepackage{graphicx}
\usepackage{color}
\usepackage{upgreek}
\usepackage{comment}
\usepackage{hyperref}
\usepackage{epstopdf}
\usepackage{caption}
\usepackage{subcaption}
\usepackage{braket}
\usepackage{bm}

\DeclareMathAlphabet{\mathbold}{U}{zeur}{b}{n}

\usepackage[dvipsnames]{xcolor}
\usepackage[labelfont=bf,font={sl,small}]{caption}

\renewcommand\[{\left[}
\renewcommand\]{\right]}

\newcommand{\eq}{\begin{equation}}
\newcommand{\eqq}{\end{equation}}

\newcommand{\bec}[1]{\mathrm{\textbf{#1}}}

\def\beq{\begin{equation}}
\def\eeq{\end{equation}}
\def\[{\begin{equation}}
\def\]{\end{equation}}

\begin{document}
\numberwithin{equation}{section}

\title{Semiclassical computation of quantum effects in multiparticle production at large $\bm{\lambda n}$
}

\author{Valentin V. Khoze}

\affiliation{Institute for Particle Physics Phenomenology, Department of Physics, Durham University,\\ 
Durham, DH1 3LE, UK}

\emailAdd{valya.khoze@durham.ac.uk}

\abstract{
We use the semiclassical formalism based on singular solutions in complex time to compute scattering rates for multiparticle production at high energies.
In a weakly coupled $\lambda \phi^4$ scalar field theory in four dimensions, we consider 
scattering processes where the number of particles $n$ in the final state approaches its maximal value $n \to E/m \gg 1$, where $m$ is the particle mass. Quantum corrections to the known tree-level amplitudes in this regime are characterised by the parameter $\lambda n$ and we show that they become large at sufficiently high multiplicities. We compute full amplitudes in the large $\lambda n$ limit on multiparticle mass thresholds using the thin-wall realisation of the singular solutions in the WKB approach.
We show that the scalar theory with spontaneous symmetry breaking, used here as a simplified model for the Higgs sector, leads to exponentially growing multi-particle rates within our regime which is likely to realise the high-energy Higgsplosion phenomenon. 
We also comment on realisation of Higgsplosion in dimensions lower than four.
}

\date{}

\preprint{IPPP/18/46}
\maketitle


\section{Introduction}
\label{sec:intro}

The aim of this paper is to present and explain the semiclassical calculation
of ${\rm few} \to n$ particle processes in the limit of ultra-high particle multiplicities $n$.
The underlying semiclassical formalism,
was originally developed by Son in Ref.~\cite{Son:1995wz}, while a first version of the calculation was presented in my earlier paper~\cite{Khoze:2017ifq}.
The present paper seeks to provide a more detailed justification of the main result and its derivation.  

\medskip

We are interested in $1^* \to n$ decay rates where $1^*$ is a virtual state created by a local operator
${\cal O}(x)$ at a point $x=0$. In high-energy scattering processes the highly virtual states $1^*$  with $Q^2 =s$ would correspond to the $s$-channel 
resonances created by the two incoming colliding particles. For example in the gluon fusion process,
$gg \to h^* \to n\times h$ the highly virtual Higgs boson $h^*$ is created by the two initial gluons before decaying into $n$ Higgs bosons in
the final state. The $1^* \to n$ decay rates we are interested in, correspond in this example to the  $h^* \to n\times h$ part of the
process.

As this paper is about proving a technical point by providing a non-perturbative calculation of the $n$-particle decay rates, 
we leave the discussion and interpretations of the resulting rates, which will turn out to be unsuppressed in the model we are considering,
to other papers and future work. The calculation that we present is aimed towards developing a theoretical foundation 
for the phenomenon of Higgsplosion proposed in \cite{Khoze:2017tjt} and further investigated in the recent papers 
\cite{Khoze:2017lft,Gainer:2017jkp,Khoze:2017uga,Khoze:2018bwa}.

 As in Refs.~\cite{Khoze:2017ifq,Khoze:2017tjt} we are interested in the scalar sector of the theory
 which for simplicity we will take to be a quantum field theory of a single real degree of freedom $h(x)$ described by the Lagrangian,
\[
{\cal L} \,= \, 
\frac{1}{2}\, \partial^\mu h \, \partial_\mu h\, -\,  \frac{\lambda}{4} \left( h^2 - v^2\right)^2
\,. \label{eq:L}
\]
The theory has a non-zero vacuum expectation value $\langle h\rangle=v$ which breaks spontaneously the $Z_2$ symmetry, and gives the mass 
$m = \sqrt{2\lambda}\,v$ to the elementary scalar particle described by the shifted field,
\[
\phi(x)=h(x)-v \,.
\]
This model can be viewed as a reduction of the SM Higgs sector in the unitary gauge to a single scalar field.
In this simplified model the scalar boson is all there is, and since all other SM-like degrees of freedom (vector bosons and fermions) are decoupled,
the scalar $h(x)$ is stable.

\bigskip

Our goal is to compute the multi-boson production rate in the large $\lambda n$ limit,
where $\lambda$ is the coupling constant and $n$ is the particle number in the final state.
On the technical side, the idea which makes this calculation possible, is to 
combine the semiclassical formalism developed by Son in Ref.~\cite{Son:1995wz} 
based on singular classical solutions  with the 
idea \cite{Khoze:2017ifq}
to search for these solutions
in the form of thin-walled singular bubbles. The thin-wall approximation has been already adopted to multiparticle production processes earlier
in Ref.~\cite{Gorsky:1993ix} in the case of standard non-singular smooth bubble configurations as in the false vacuum decay.
We will instead tie the appearance of the semiclassical configurations with singular thin-wall surfaces to the requirements of the 
semiclassical approach  Ref.~\cite{Son:1995wz}. 

In the scattering processes at very high energies, production of large numbers of particles in the final state 
becomes possible.
These processes
were studied in some detail in the  literature and we refer the reader to papers
\cite{Cornwall:1990hh,Goldberg:1990qk,Brown:1992ay,Argyres:1992np,Voloshin:1992rr,Voloshin:1992nu,
Libanov:1994ug,Gorsky:1993ix,Libanov:1997nt,Khoze:2014zha,Khoze:2014kka,Jaeckel:2014lya,Khoze:2015yba,Degrande:2016oan,Jaeckel:2018ipq}
and references therein.  

This paper is organised as follows. In section \ref{sec:class} we briefly recall the known results for the multiparticle scattering rates 
obtained in perturbation theory at tree-level, before proceeding with  the non-perturbative calculation in the main body of the paper.
In section~\ref{sec:WKB1} we will summarise the semiclassical approach of Son as a series of steps needed to identify
the saddle-point solution in Minkowski space. In section~\ref{sec:WKB2}, still following~\cite{Son:1995wz}, we simplify and refine this prescription 
as the extremization over singular surfaces approach in complex time. The resulting set-up is ideal for using the thin-wall approach which we develop 
is sections~\ref{sec:son_loops}
and~\ref{sec:thin_w}. In particular, in section~\ref{sec:son_loops} we will recover tree-level results familiar from section \ref{sec:class} along with
the prescription for computing the quantum corrections. These quantum contributions to the multi-article rates are computed in section~\ref{sec:thin_w} using the thin-walled 
singular classical solutions. In section~\ref{sec:lowd} we consider multiparticle processes in 3 dimensions and provide a successful test for the semiclassical results.
Finally,
we present our conclusions in section~\ref{sec:concl}.

\newpage
 
\section{Simple classical solutions and tree-level amplitudes at threshold}
\label{sec:class}

The purpose of this paper is to compute the amplitudes and the corresponding probabilistic rates for processes involving multiparticle final states 
in the large $\lambda n$ limit non-perturbatively -- i.e. using a semiclassical approach with no reference to perturbation theory 
and without artificially separating the result into a tree-level and a `quantum corrections' contributions. Their entire combined contribution should emerge from the unified semiclassical algorithm. But to first set the scene for such a computation we need to recall the known properties 
of the tree-level amplitudes and their relation with certain classical solutions. This is the aim of this section.

Thus, we start here with tree-level $n$-point scattering amplitudes 
computed on the $n$-particle mass thresholds. This is the kinematics regime where all $n$
final state particles are produced at rest. These amplitudes for all $n$ are conveniently assembled 
into a single object -- the amplitude generating function -- which at tree-level is described by 
a particular solution of the Euler-Lagrange equations.
The classical solution which provides the generating function of tree-level amplitudes on multi-particle
mass thresholds in the model \eqref{eq:L} is given by~\cite{Brown:1992ay},
\[
h_{0} (z_0;t) \,=\, v\, \left(\frac{1+z_0\, e^{im t}/(2v)}{1-z_0\, e^{im t}/(2v)}\right)\,, \quad m=\sqrt{2\lambda} v\,,
\label{clas_sol}
\]
and where $z_0$ is an auxiliary variable. It is easy to check with the direct substitution that the
expression in \eqref{clas_sol} does indeed satisfy the Euler-Lagrange equation resulting from our theory
Lagrangian \eqref{eq:L} for any value  of the $z_0$ parameter. It then follows 
that all $1^*\to n$ tree-level scattering amplitudes on the $n$-particle mass thresholds are given by the differentiation 
of $h_{0} (z_0;t)$ with respect to $z_0$,
\[
{\cal A}_{1\to n}\,=\, \langle n|S \phi(0)| 0\rangle \,=\,\, \left.\left(\frac{\partial}{\partial z_0}\right)^n h_{0} \,\right|_{z_0=0}
 \label{eq:A5}
\]
The classical solution in \eqref{clas_sol} is uniquely specified by requiring that it is a holomorphic function of the complex variable 
$z(t) = z_0\, e^{im t}$,
\[
h_{0} (z) \,=\,  v\,+\, 2v\,\sum_{n=1}^{\infty} \left(\frac{z}{2v}\right)^n
\,, \quad z=z(t) = z_0\, e^{im t}\,,
\label{gen-funh2}
\]
so that the amplitudes in \eqref{eq:A5} are given by the coefficients of the Taylor expansion in \eqref{gen-funh2} 
times $n!$ from differentiating $n$ times over $z$,
\[
{\cal A}_{1\to n}\,=\, 
\left.\left(\frac{\partial}{\partial z}\right)^n h_{0} (z) \,\right|_{z=0} \,=\, n!\, \left(\frac{1}{2v}\right)^{{n-1}}\,=\, 
n!\, \left(\frac{\lambda}{2m^2}\right)^{{\frac{n-1}{2}}}
\,.
\label{eq:ampln2}
\]
These formulae and the characteristic factorial growth of $n$-particle amplitudes,  ${\cal A}_{n} \sim \lambda^{n/2} n!$, 
form the essence of the elegant formalism pioneered by Brown in Ref.~\cite{Brown:1992ay} that is 
based on solving classical equations of motion and bypasses the summation over individual Feynman diagrams. 
In the following sections we will see how these (and also more general solutions describing full quantum processes) emerge
from the semiclassical approach of \cite{Son:1995wz} which we shall follow.

 \begin{figure*}[t]
\begin{center}
\includegraphics[width=0.8\textwidth]{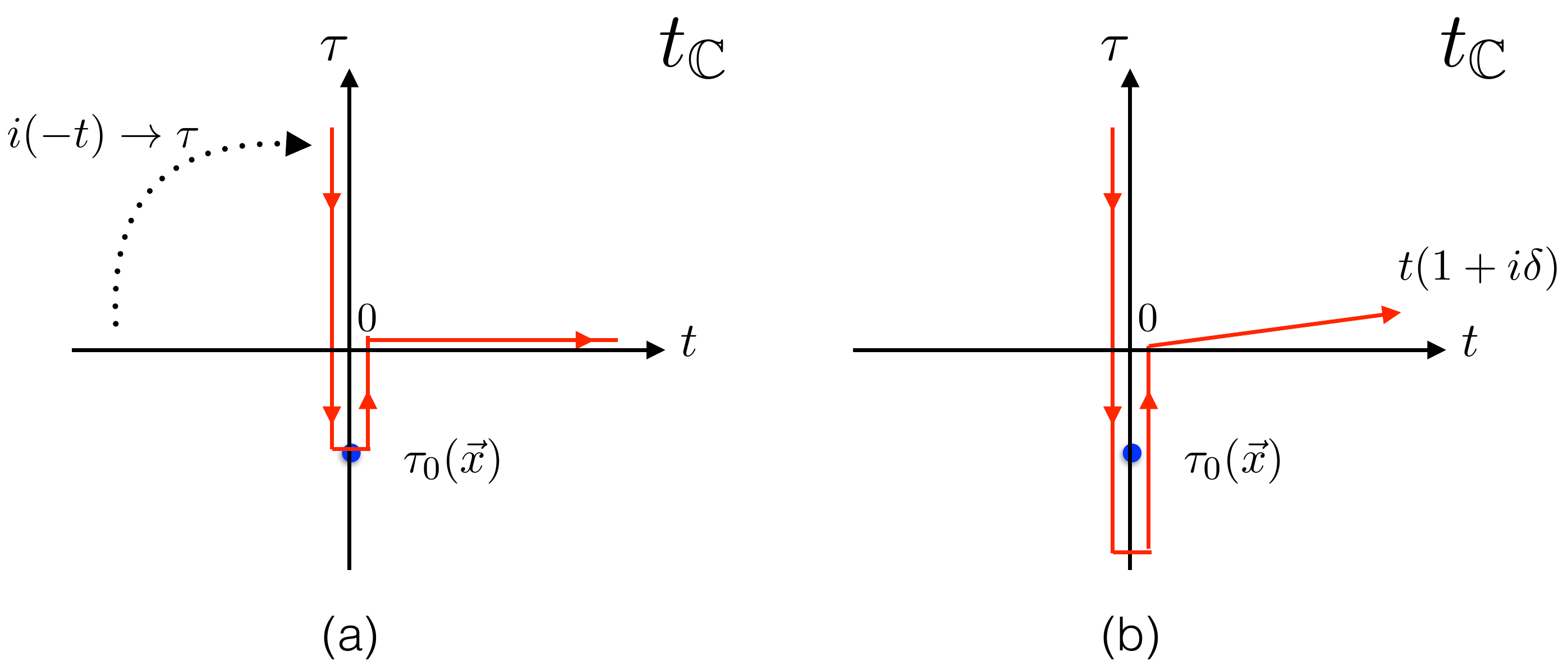}
\end{center}
\vskip-.5cm
\caption{Time evolution contour on the complex time plane $t_{\mathbb{C}}$.  {\bf Plot (a)} shows the contour obtained after deforming the 
the evolution along the real time axis $-\infty<t<+\infty$ where the early-time ray $-\infty <t<0$ is rotated by $\pi/2$ into the ray along vertical axis,
$\infty>\tau>\tau_0({\bf x})$ and ending at the singularity surface of the solution $\tau_0({\bf x})$.
{\bf Plot (b)} shows a refinement of this contour: (1) rather than touching the singularity, the contour surrounds it; (2) at the
late time boundary condition, the contour approaches $t\to +\infty$ along the ray with an infinitesimally small positive angle $\delta$ to the real time axis.}
\label{fig:contour}
\end{figure*}

We note that the classical solution \eqref{gen-funh2} is complex-valued. This is in spite the fact that we are working with the
real-valued scalar field theory model \eqref{eq:L}. The classical solution $h_0$ that generates tree-level amplitudes via \eqref{eq:ampln2}
does not have to be real, in fact it is manifestly complex (in real time) and this is a consequence of the fact that this solution will emerge
as an extremum of the action in the path integral using the steepest descend method. 
In this case the integration contours in path integrals are deformed to enable them to pass through extrema (or encircle singularities) 
that are generically complex-valued.

We will be working with classical solutions and other field configurations that depend on 
the complexified time $t_\mathbb{C}$. Hence we promote the real time variable $t$ into the variable $t_\mathbb{C}$ that takes 
values on the complex time plane,
\[
t\,  \longrightarrow\, t_{\mathbb{C}} = t+i\tau\,,
\label{eq:tC}
\]
where $t$ and $\tau$ are real valued. We will use the deformation the time-evolution contour from the real time axis $-\infty <t<+\infty$
to the contour in the complex $t_{\mathbb{C}}$ plane depicted in Fig.~\ref{fig:contour}
in such a way that the initial time $t=-\infty$ maps on the imaginary time ${\rm Im}\, t_{\mathbb{C}} \,=\,\tau=+\infty$. This corresponds to the 
$(- t) \times e^{i\pi} = \tau$ rotation, 
\begin{eqnarray}
{\rm at \, early \, times,\,} -\infty<t<0: &&\quad  t \to i\tau
\label{eq:ourAC}
\end{eqnarray}
We also note that $\tau$ corresponds to minus the Euclidean time $t_{\rm Eucl}$ defined by the standard Wick rotation via $t\to -i t_{\rm Eucl}$.

Expressed as the function of the complexified time variable $t_{\mathbb{C}}$, the classical solution \eqref{clas_sol} reads,
\[
h_{0} (t_{\mathbb{C}}) \,=\, v\, \left(\frac{ 1\,+\,e^{im (t_{\mathbb{C}}-i\tau_\infty)}}
{1\,-\, e^{im (t_{\mathbb{C}}-i\tau_\infty)}}\right)\,,
\label{clas_sol2}
\]
where  $\tau_\infty$ a constant,
\[ 
\tau_\infty := \frac{1}{m} \log \left(\frac{z_0}{2v}\right)\,
\]
it parameterises the location (or the centre)
of the solution in imaginary time. If the time-evolution contour of the solution in the $t_{\mathbb{C}}$ plane is along the 
the imaginary time with the real time $t=0$, the field configuration \eqref{clas_sol2} becomes real-valued,
\[
h_{0} (\tau) \,=\, v\, \left(\frac{ 1\,+\,e^{-m (\tau-\tau_\infty)}}
{1\,-\, e^{-m (\tau-\tau_\infty)}}\right)\,,
\label{clas_sol22}
\]
and singular at $\tau=\tau_\infty$. 

Having already noted that the solution is complex-valued we note another important feature of the solution \eqref{gen-funh2} 
that is for the forthcoming semiclassical analysis, 
namely that the configuration $h_0$ is singular in imaginary time, in particular at $\tau = \tau_{\infty}$ when $t=0$.

 \begin{figure*}[t]
\begin{center}
\includegraphics[width=0.55\textwidth]{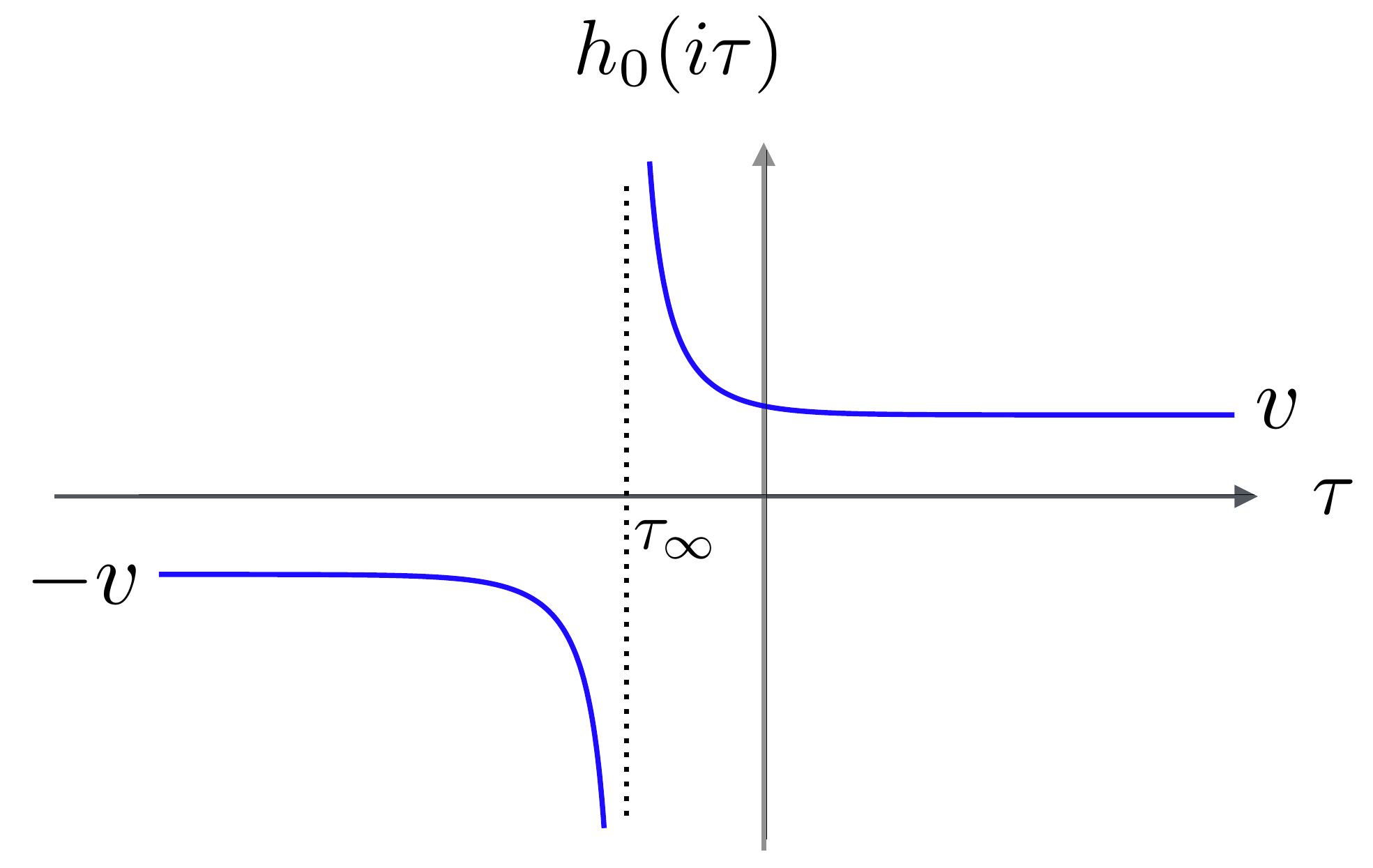}
\end{center}
\vskip-.5cm
\caption{Singular classical solution \eqref{clas_sol22} uniform in space: flat domain wall located at $\tau_\infty$ in the imaginary time.}
\label{fig:kink}
\end{figure*}

The expression on the right hand side of \eqref{clas_sol22} has an obvious interpretation 
in terms of a singular domain wall located at $\tau=\tau_\infty$ that separates two domains of the field $h(\tau, {\bf x})$ as shown in Fig.~\ref{fig:kink}
The domain on the right of the wall 
$\tau \gg \tau_\infty$ has $h = +v$,
and the domain on the left of the wall, $\tau \ll \tau_\infty $, is characterised by $h = -v$.
The field configuration is singular at the position of the wall, $\tau=\tau_\infty$, for all values of ${\bf x}$, i.e. the singularity surface is 
flat (or uniform in space). The thickness of the wall is set by the inverse mass $1/m$.

The field configuration \eqref{clas_sol22} can be used to compute
 the surface tension of the domain wall.  The surface tension is defined as the Euclidean action 
 computed on \eqref{clas_sol22} per unit area of the 3-dimensional surface 
 $\tau_0({\bf x})=\tau_\infty$.
Since the  $\tau_\infty$ surface is uniform in space, the surface tension is given by the integral,
\[
\mu\,=\, \int_{-\infty+i\epsilon}^{+\infty+i\epsilon} d\tau \left( \frac{1}{2} \left(\frac{d h}{d\tau}\right)^2 +
 \frac{\lambda}{4} \left( h^2 - v^2\right)^2 \right) \,=\,  \frac{m^3}{3 \lambda}\,,
 \label{eq:mu22}
\]
This integral is finite for the contour along $\tau$ shifted by $\pm i \epsilon$; the rational for this procedure will be explained 
in section~\ref{sec:thin_w} {\it cf.} Eq.~\eqref{eq:mu}.

In the following section we will summarise the results of the semiclassical formalism for computing probability rates of $1^* \to n$ processes 
in which the complex-valued singular configurations of the type \eqref{clas_sol2} appear naturally as the solution of the boundary value problem.

\section{The semiclassical formalism of Son}
\label{sec:WKB1}

Motivated in part by the Landau formulation of the WKB approach in the non-relativistic quantum mechanics \cite{Landau,Landau2},
D.~T.~Son developed in Ref.~\cite{Son:1995wz} a semiclassical formalism for computing multi-particle
cross-sections in a quantum field theory. It relies on functional integrals in the coherent state representation to 
specify the initial and final states as the boundary conditions at early and late times. The functional integrals are then evaluated using the
steepest descent method with the dominant field configurations and other relevant parameters taking in general complex values. The 
complex-valued saddle points (local minima in our model) and the presence of singularities in the solutions of the boundary value problem 
are the essential characteristics of the Landau-WKB and the Son's approach in quantum field theory.\footnote{Earlier work 
on generalisations of the Landau-WKB formalism to
problems with many degrees of freedom includes 
Refs.~\cite{IP,Voloshin:1990mz,Khlebnikov:1992af,Diakonov:1993ha} and in section 4 of the review \cite{Libanov:1997nt}.}
 
In this section we will list the main steps that specify the steepest descent solution of the boundary value problem 
in the formalism of Son. These steps follow directly from the construction in ~\cite{Son:1995wz},
and for the convenience of the reader in the {\tt Appendix A} we provide additional comments on the algorithm.
No prior familiarity with the formalism in \cite{Son:1995wz} is required to follow the algorithm for finding the solution,
however a pedagogical overview of Ref.~\cite{Son:1995wz} is beyond the scope of this paper; this task is 
postponed to a separate work~\cite{Khoze:2018mey}.

The central quantity is the dimensionless probability rate ${\cal R}_n(E)$ for a local operator ${\cal O}(x)$ at a point $x=0$
to create $n$ particles of total energy $E$ from the vacuum. It is given by \cite{Son:1995wz},
\[
{\cal R}_n(E)\,=\, \int d\Phi_n\, \langle 0| \,{\cal O}^\dagger \, S^\dagger\, P_E |n\rangle 
\langle n|\, P_E\, S\, {\cal O}\, |0\rangle\,,
\label{eq:RnE1}
\]
where the matrix element involves the operator ${\cal O}$  between the vacuum state $ |0\rangle$ and the $n$-particle state
of fixed energy $\langle n|\, P_E $ (here $P_E$ is the projection operator on states with fixed energy $E$), along with
the $S$ matrix to evolve between the initial and finial times. The matrix element is squared and integrated over the 
$n$-particle Lorentz-invariant phase space.
The local operator ${\cal O}$ appearing in the matrix elements in \eqref{eq:RnE1} is conventionally \cite{Son:1995wz}
in the form
\[ 
{\cal O} \,=\,j^{-1}\, {e^{j\,(h(0)-v)}}\,=\, j^{-1} \, e^{j \phi(0)}\,,
\label{eq:opdef}
\]
where $j$ is a constant, and the limit $j\to 0$ is taken in the computation of the probability rates \eqref{eq:RnE1} 
to select the single particle initial state $\langle 0| \phi(0)$. 

The cross-sections for few to many particles, $\sigma_{{\rm few}\to n} (E)$ as well as multi-particle partial decay 
rates $\Gamma_n(E)$ of a single particle state $X \to n\times h$, are determined by the exponential factor for
${\cal R}_n(E)$ in \eqref{eq:RnE1} times a non-exponential prefactor of appropriate dimensionality which is 
of no interest in a semiclassical approximation. 

In the construction of~\cite{Son:1995wz} the expression on the right hand side of  \eqref{eq:RnE1} 
is represented as a functional integral, which is subsequently computed in the steepest descent approximation
for all integration variables. 
The steepest descent method relies on having a single large parameter in front of all terms in the exponent. This parameter is
the inverse coupling constant $1/\lambda \gg1$ in the weak-coupling limit of the theory. The final state particle number 
$n=\lambda n /\lambda$ is $ \sim 1/\lambda$ for $\lambda n={\rm fixed}.$ 
Thus the steepest descent method is justified in the
double-scaling weak-coupling and large-$n$ semiclassical limit:
\[ \lambda \to 0\,, \quad n\to \infty\,, \quad {\rm with}\quad
\lambda n = {\rm fixed}\,, \quad \varepsilon ={\rm fixed} \,.
\label{eq:limit}
\]
Here $\varepsilon$ denotes the average kinetic energy per particle per mass in the final state,
\[ \varepsilon \,=\, (E-nm)/(nm)\,.
\]
Holding $\varepsilon$ fixed implies that in the large-$n$ limit we are raising the total energy linearly with $n$.
The semiclassical result for the rate has the characteristic exponential form \cite{Son:1995wz},
\[
{\cal R}_n(E)\,\simeq \, \exp \left[ W(E,n)\right],
\label{eq:ReW}
\]
where 
\[ W(E,n) \,\equiv\, \frac{1}{\lambda}\, {\cal F}(\lambda n, \varepsilon)\,=\,\,
ET \,-\, n\theta \,-\, 2{\rm Im} S[h]\,
\label{eq:Wdef}
\]
$S$ is the action on the complex-valued field solution and $T$ and $\theta$ are the auxiliary parameters 
that will be specified momentarily.

\bigskip

\noindent The algorithm~\cite{Son:1995wz} to find the
saddle-point configuration on which to compute the semiclassical rate ${\cal R}_n(E)$ is as follows:


\begin{enumerate}
\item Solve the classical equation without the source-term,
\[
\frac{\delta S}{\delta h(x)}\,=\,0\,,
\label{eq:al1}
\]
by finding a complex-valued solution $h(x)$ with a point-like singularity at the origin
$x^\mu=0$ and regular everywhere else in Minkowski space. The singularity at the origin is selected by the location of the operator ${\cal O}(x=0)$.

\item Impose the initial and final-time boundary conditions,
\begin{eqnarray}
\lim_{t\to - \infty}\,h(x)  &=& v\,+\, 
\int \frac{d^3k}{(2\pi)^{3/2}} \frac{1}{\sqrt{2\omega_{\bf k}}}\,\, a^*_{\bf k}\, e^{ik_\mu x^\mu} \,
\label{eq:al2}
\\
\lim_{t\to + \infty}\,h(x) &=& v\,+\, 
\int \frac{d^3k}{(2\pi)^{3/2}} \frac{1}{\sqrt{2\omega_{\bf k}}}\left( b_{\bf k}\,e^{\omega_{\bf k}T-\theta}
\, e^{-ik_\mu x^\mu}\,+\, b^*_{\bf k}\, e^{ik_\mu x^\mu}\right)\,.
\label{eq:al3}
\end{eqnarray}

\item Compute the energy and the particle number using the 
$t\to +\infty$ asymptotics of $h(x)$,
\[
E \,=\, \int d^3 k \,\, \omega_{\bf k}\, b_{\bf k}^\dagger \, b^{}_{\bf k}\, e^{\omega_{\bf k}T-\theta}
\,, \qquad
n \,=\, \int d^3 k \,\, b_{\bf k}^\dagger \, b^{}_{\bf k}\, e^{\omega_{\bf k}T-\theta}\,.
\label{eq:al4}
\]
At $t\to -\infty$ the energy and the particle number  are vanishing. 
The energy is conserved by regular solutions and changes discontinuously from $0$ to $E$ 
at the singularity at $t=0$.

 \item Eliminate the $T$ and $\theta$ parameters in favour of $E$ and $n$ using the expressions above.
 Finally, compute the function $W(E,n)$
  \[ 
W(E,n)  \,=\,
ET \,-\, n\theta \,-\, 2{\rm Im} S[h]
\label{eq:al5}
\]
on the set $\{h(x), T, \theta\}$ and compute the semiclassical rate ${\cal R}_n(E) \,=\, \exp \left[ W(E,n)\right]$.
\end{enumerate}

\bigskip

\noindent To implement this programme one starts with the specified expressions \eqref{eq:al2} and \eqref{eq:al3}
for $h(x)$ at the $t\to \pm \infty$ boundaries and classically
evolves them by solving the equation of motion into the region of
finite $t$. We thus have two trial functions, one at $t<0$ and the 
second at $t>0$  which we would like to match at $t=0$.
The field configuration at $t<0$ is given by a regular classical solution $h_1(t,{\bf x})$ which 
satisfies the initial time boundary condition with the Fourier coefficient functions $a^*_{\bf k}$.
The second trial function, $h_2(t,{\bf x})$, is a regular classical solution on the Minkowski half-plane $t>0$
which is evolved from the final-time boundary condition with the coefficient functions $b_{\bf k}\,e^{\omega_{\bf k}T-\theta}$
and $b^*_{\bf k}$.
One then contemplates scanning over the space of the functions $a_{\bf k}$ and $b_{\bf k}$ to achieve
the matching at $t=0$  between the two branches $h_1$ and $h_2$ of the solution, $h_1({\bf x})=h_2({\bf x})$, 
and all of its time derivatives for all values of ${\bf x}\neq 0$.
The only allowed singularity of the full solution is point-like, and located at the origin $t=0={\bf x}$.

A practical difficulty in implementing the matching between $h_1$ and $h_2$ is that
 $h_1(x)$ should be equal to $h_2(x)$ on the entire hyperplane $(t=0, {\bf x})$ 
with the exception of the single point $t=0= {\bf x}$.
This technical difficulty can be bypassed following \cite{Son:1995wz}, by analytically continuing to complex time as we will explain in the following section.

\section{Refining the method in complex time $t_\mathbb{C}$}
\label{sec:WKB2}

In Minkowski space-time $x^\mu=(t,{\bf x})$ the desired solution $h(x)$ should contain a point-like singularity at the origin $x=0$, 
and be regular everywhere else. In the 
Euclidean space-time, $(\tau,{\bf x})$, however, such a solution will in general be singular on a 3-dimensional hypersurface
$\tau = \tau_0({\bf x})$ located at $t=0$. 

To illustrate this point consider the already familiar from section \ref{sec:class} classical solution \eqref{clas_sol2}.
We now modify this field configuration by replacing the collective coordinate parameter $\tau_\infty$ by a
function $\tau_0({\bf x})$ that is no longer uniform in space, but interpolates between $0$ at ${\bf x}=0$ and a constant  
$\tau_\infty$ at $|{\bf x}| \to \infty.$ 
The configuration 
\[
h_{0} (t_{\mathbb{C}};{\bf x}) \,=\, v\, \left(\frac{ 1\,+\,e^{im (t_{\mathbb{C}}-i\tau_0({\bf x}))}}
{1\,-\, e^{im (t_{\mathbb{C}}-i\tau_0({\bf x}))}}\right)\,,
\label{clas_sol2new}
\]
deviates from an exact solution of equations of motion by terms involving derivatives of $\tau_0({\bf x})$ and requires additional 
corrections on the right hand side, but for a slowly 
varying $\tau_0$ it is a good trial function to expand around and use in a variational principle. It then immediately follows 
that in Minkowski spacetime where $t_{\mathbb{C}}=t$ is real, the field configuration \eqref{clas_sol2new} is singular
at the point $t=0={\bf x}$, while in complex time, it is singular on the surface located at 
$t_{\mathbb{C}}= 0 + i \tau_0({\bf x})$ spanned by the 3-dimensional variable ${\bf x}$.

 \begin{figure*}[t]
\begin{center}
\includegraphics[width=0.85\textwidth]{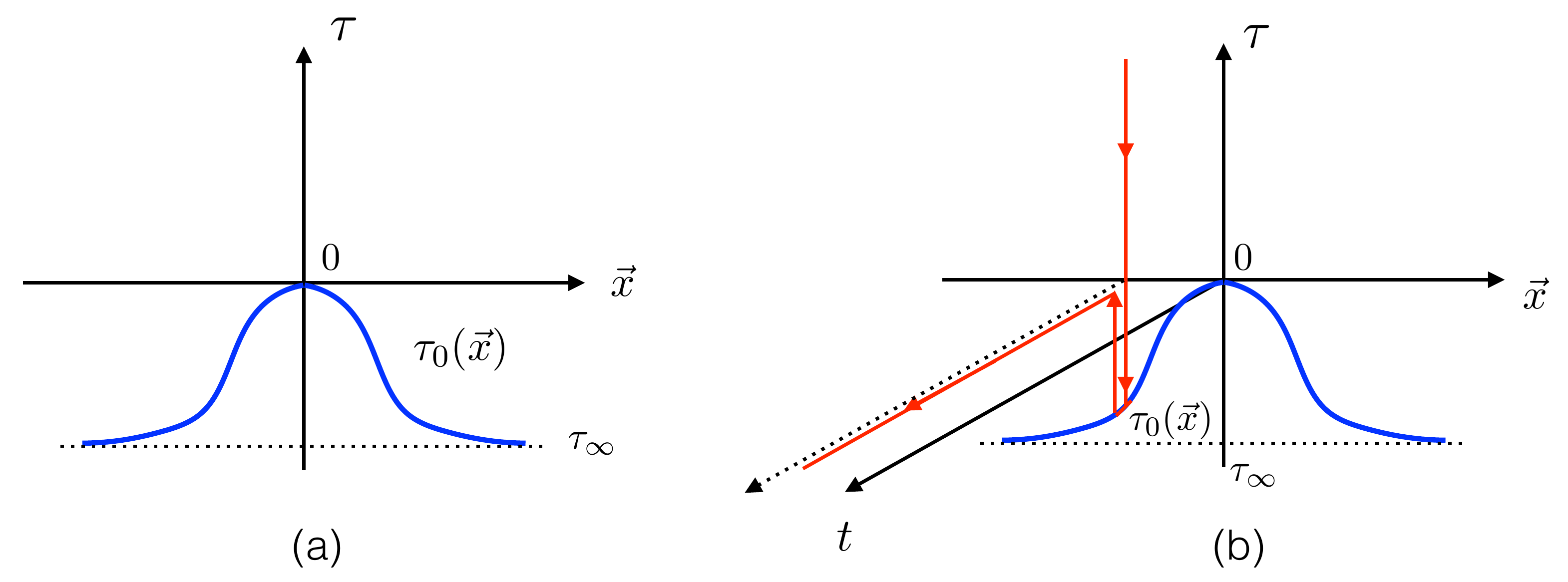}
\end{center}
\vskip-.5cm
\caption{{\bf Plot (a)} shows the shape of the singularity surface $\tau_0({\bf x})$ of the field configuration $h(x)$ on the imaginary time hyperplane $(\tau,{\bf x})$.
{\bf Plot (b)} shows the time evolution contour of Fig.~\ref{fig:contour}~(a) in the coordinate system $(t,\tau;{\bf x})$.}
\label{fig:tau}
\end{figure*}
\bigskip

\noindent We now describe the extremization procedure for finding the solution to the boundary value problem 
in complexified time $t_{\mathbb{C}}= t + i \tau$, following \cite{Son:1995wz}:

\begin{enumerate}
\item Select a trial singularity surface located at $\tau=\tau_0({\bf x})$.
The surface profile $\tau_0({\bf x})$ is an ${\cal O}(3)$ symmetric function of ${\bf x}$ and is given by a local 
deformation of the flat singularity domain wall at $\tau_\infty$ with the single maximum touching the origin $(\tau, {\bf x})=0$
as shown in Fig.~\ref{fig:tau}~(a).
In Minkowski space the singularity is point-like at $t=0=\tau$ and ${\bf x}=0$ as required.

\item Deform the time evolution contour specifying the paths in the Feynman path integral to follow the contour 
on the complex plane $(t,\tau)$,
\[
\, \left[(0, \infty) \to (0,\tau_0({\bf x}))\right] \, \oplus\,  \left[(0,\tau_0({\bf x}))\to (0,0)\right] \, \oplus\, \left[(0,0) \to (\infty,0)\right] 
\,,
\label{eq:contour}
\]
as shown in Figs.~\ref{fig:tau}~(b) and ~\ref{fig:contour}~(a).
More precisely, in order to be able to linearise the late time asymptotics of the solution, as in \eqref{eq.E3} below,
we should make the final third segment of the contour in \eqref{eq:contour} to have an infinitesimal positive angle 
w.r.t. the real time axis, i.e. $t(1+\delta)$ for $0\le t< +\infty$ with $\delta=0_+$.

\item Find a classical trajectory $h_1(\tau,{\bf x})$ on the first segment, $ +\infty > \tau > \tau_0({\bf x})$, of the contour
\eqref{eq:contour} that satisfies the initial time (vanishing) boundary condition \eqref{eq:al2},
\[
\lim_{\tau\to + \infty}\, h_1(\tau,{\bf x})  \,-\, v 
\,\,\to \,\, 0\,,
\label{eq.E1}
\]
and becomes singular as $\tau \to \tau_0({\bf x})$ so that\footnote{One can always assume a regularisation procedure that keeps $\Phi_0$ finite 
at intermediate stages of the calculation, i.e. before taking the limit of the operator source $j\to 0$.}
 $h_1(\tau,{\bf x})|_{\tau\to \tau_0({\bf x})} \,\equiv\, \Phi_0 \,\to\, \infty$.
 
\item Find another classical solution $h_2 (\tau,{\bf x})$ on the remaining part of the contour \eqref{eq:al2},
that at $\tau \to \tau_0({\bf x})$ is singular and matches with $h_1$,
\[
h_2 (\tau_0,{\bf x}) \,= \, h_1(\tau_0,{\bf x}) \,=\, \Phi_0 \,\to\, \infty\,,
\label{eq.E2}
\]
and also satisfies the final time 
boundary condition \eqref{eq:al3},
\[
\lim_{t\to + \infty}\,h_2(t,{\bf x}) \,-\, v  \,=\, 
\int \frac{d^3k}{(2\pi)^{3/2}} \frac{1}{\sqrt{2\omega_{\bf k}}}\left( b_{\bf k}\,e^{\omega_{\bf k}T-\theta}
\, e^{-ik_\mu x^\mu}\,+\, b^*_{\bf k}\, e^{ik_\mu x^\mu}\right)\,.
\label{eq.E3}
\]
The two functions $h_1(\tau,{\bf x})$ and $h_2(\tau,{\bf x})$ can be viewed as the two branches of a trial configuration $h(x)$.
The action of $h(x)$ along our complex-time contour is the sum of the action integrals\footnote{As usual,
 ${\cal L}_{\rm Eucl}[h] \,:= \, {\rm kinetic}\,+\, {\rm potential}\,=\,
\frac{1}{2}\, (\partial_\mu h)^2\, +\,  \frac{\lambda}{4} \left( h^2 - v^2\right)^2$.} of $h_1(\tau,{\bf x})$ and $h_2(t,{\bf x})$ 
on the parts of the contour,
\[
iS[h]\,=\,\int  d^3 x \left( \int_{+\infty}^{\tau_0({\bf x})} d\tau \, {\cal L}_{\rm Eucl}(h_1) \,+\,
\int_{\tau_0({\bf x})}^0d\tau \, {\cal L}_{\rm Eucl}(h_2) \,+\,i 
 \int_{0}^\infty dt \, {\cal L}(h_2) \right)
 \label{eq:Se}
\]

\item Up to this point we have not imposed the matching conditions on the
{\it derivatives} of  $h_1(\tau,{\bf x})$ and $h_2(\tau,{\bf x})$ at the singularity surface at $\tau_0$. A priori, the normal
derivatives to the surface will be different,  $\partial_n (h_1\,-\,h_2)|_{\tau\to \tau_0({\bf x})}\, \neq0$,
and the equation of motion \eqref{eq:al1} will not be satisfied at the matching surface $\tau_0({\bf x})$. 
For the combined configuration $h(x)$ to solve the classical equation \eqref{eq:al1} everywhere, 
including the $\tau_0$ surface, one simply needs 
to extremize the action integral  \eqref{eq:Se} over all singularity surfaces $\tau=\tau_0({\bf x})$
containing the point $t=0={\bf x}$.

\item Finally, determine the semiclassical rate by evaluating
\[ 
W(E,n)  \,=\,
ET \,-\, n\theta \,-\, 2{\rm Im} \,S [h]
\label{eq:al52}
\]
on the extremum, using \eqref{eq:Se} for the action,  and expressions for $T$ and $\theta$ in terms of of $E$ and $n$ found from \eqref{eq:al4} as before.
The imaginary part of the Minkowski action in \eqref{eq:Se}, \eqref{eq:al52} is the same as the real part of the Euclidean action, 
$iS :=-S_{\rm Eucl}$ and
$2{\rm Im} \,S \,=\, -iS+iS^* \,=\, 2{\rm Re} \,S_{\rm Eucl}.$

\end{enumerate}

\noindent This is the general outcome of the semiclassical construction  of  Ref.~\cite{Son:1995wz}. 
One starts with the two individual solutions satisfying the boundary conditions \eqref{eq.E1}-\eqref{eq.E3} and then varies over the
profiles of the singular matching surface $\tau_0({\bf x})$ to find an extremum 
of the imaginary part of the action  \eqref{eq:Se}. On the extremal surface not only the field configurations, but also their 
normal derivatives match $\partial_n (h_1\,-\,h_2)=0$ at all ${\bf x}$ except ${\bf x}=0.$ This implies that $h_1=h_2$ on the entire slice of the 
spacetime where they are both defined, i.e. for $\tau$ in the interval $[0,\tau_0]$, except at the point at the origin. Restricting to the
Minkowski space slice, i.e. at $\tau=0$, this implies $h_1(0,{\bf x})=h_2(0,{\bf x})$, as it should be. It does not mean however that the real part of the action in \eqref{eq:Se} vanishes, as the sum of the first two integrals can be viewed as encircling the singularity of the solution at $\tau_0$.

In summary, the highly non-trivial problem of searching for the appropriate singular field solutions $h(x)$ is reduced to a geometrical problem --
extremization over the surface shapes $\tau_0({\bf x})$ and accounting for the appropriate boundary conditions \eqref{eq.E1}-\eqref{eq.E3}.
This formulation of the problem is now well-suited for using the thin-wall 
approximation that will be described in section~\ref{sec:thin_w} and will allow us to address the previously unexplored in \cite{Son:1995wz} regime
at large values of $\lambda n$ where quantum non-perturbative effects are large. 

We proceed with the practical implementation of the steps 1.-6. for the model \eqref{eq:L} in the following two sections.

\medskip
\section{Computing the rate: setting the scene}
\label{sec:son_loops}
\medskip

In this section we will specify and solve the boundary conditions in \eqref{eq.E1}, \eqref{eq.E3} at the initial and final times, 
deriving the coefficient functions $b^*_{\bf k}$  and $b_{\bf k}\,e^{\omega_{\bf k}T-\theta}$ 
in \eqref{eq.E3}. We will then determine the $T$ and $\theta$ parameters 
and compute the general expression for the exponent of the rate $W(E,n)$ in \eqref{eq:al5}.

In the limit $\varepsilon= 0$, the scattering amplitude is on the multiparticle threshold, the final state
momenta are vanishing and one would naively assume that the classical solution describing this limit is uniform in space. 
This is correct for the tree-level solution but not for the solution incorporating quantum effects. In the latter case, the correct
and less restrictive assumption is that the presence of the singularity at $x=0$ deforms the flat surface of singularities near its location,
as shown in Fig.~\ref{fig:tau}.
From now on we will concentrate on the physical case where $\varepsilon$ is non-vanishing and
non-relativistic, $0< \varepsilon \ll 1$. At the same time, the parameter $\lambda n$ is held fixed and arbitrary. It will ultimately be taken to be large. 

The initial-time boundary condition \eqref{eq.E1} dictates that the solution $h_1(t_{\mathbb{C}}=i\tau,\bec{x})-v $ must vanish with exponential accuracy 
as $e^{-m\tau}$ in the limit $\tau \to \infty$. The final-time boundary condition \eqref{eq.E3} of the finite-energy solution $h_2 (x)$ requires the solution to be singular on the
singularity surface $\tau_0(\bec{x})$.
Following Son, without loss of generality, we can search for $h_2$ in the form,
\begin{eqnarray}
h_2(t_{\mathbb{C}},\bec{x}) 
&=& v\, \left(\frac{ 1\,+\,e^{im (t_{\mathbb{C}}-i\tau_\infty)}}
{1\,-\, e^{im (t_{\mathbb{C}}-i\tau_\infty)}}\right)\,+\, \tilde{\phi}(t_{\mathbb{C}},\bec{x})
\,.
\label{clas_sol2A}
\end{eqnarray}
The first term on the right-hand side is an $\bec{x}$-independent field configuration  $h_0(t_{\mathbb{C}})$.
It is an exact classical solution 
\eqref{clas_sol2} with the surface of singularities at $ t_{\mathbb{C}} = i \tau_{\infty}$, which is a $3d$ plane in ${\bf x}$,
as shown in Fig.~\ref{fig:kink}.
The second term, $\tilde{\phi}(t_{\mathbb{C}},\bec{x})$,
describes the deviation of the singular surface from the $\tau_{\infty}$-plane. 
This deviation, $\tau_0({\bf x}) -\tau_\infty$, is locally non-trivial around $\bec{x}=0$ and vanishes at 
$\bec{x}\to \infty.$ There is no loss of generality in \eqref{clas_sol2A} because the configuration $\tilde{\phi}(t_{\mathbb{C}},\bec{x})$
is so far completely unconstrained. 

Now we can start imposing the boundary conditions  \eqref{eq.E3} at $t\to +\infty$ on the expression \eqref{clas_sol2A}.
On the final segment of the time evolution contour, $t (1+i\delta_+)$ as $t\to +\infty$, the first term in \eqref{clas_sol2A} can be Taylor-expanded 
in powers of $e^{im t(1+i\delta_+)}$
and linearised thanks to $\delta_+$ being positive, giving,
\[
\lim_{t\to + \infty} h_0(x) \,-\, v \, =\, 
2v \, e^{m\tau_\infty}\, e^{imt} \,.
\label{eq.Eh0}
\]
For the second term in \eqref{clas_sol2A} we write the general expression involving the positive-frequency and the negative frequency 
components in the Fourier transform,
\[
\lim_{t\to + \infty}\, \tilde{\phi}( t, {\bf k})
  \,=\, 
\frac{1}{\sqrt{2\omega_{\bf k}}}\left(f_{\bf k}\, e^{-i\omega_{\bf k} t}\,+\, g_{-\bf k}\, e^{i\omega_{\bf k}t}\right)\,.
\label{eq.Etilphi}
\]

We will now show that for the solution in the non-relativistic limit, $\epsilon \ll 1$, the boundary conditions \eqref{eq.E3}
will require that $g_{-\bf k} = 0$ and will also impose a constraint on the coefficient function $f_{\bf k}$, so that,
\begin{eqnarray}
g_{-\bf k} &=& 0\,, \label{eq:gk0} \\
f_{\bf k=0} &=& \frac{n \sqrt{\lambda}}{2 \pi m)^{3/2}} \, e^{-m\tau_\infty}\,. \label{eq:fk0}
\end{eqnarray}

To derive \eqref{eq:gk0}-\eqref{eq:fk0} we proceed by combining the asymptotics \eqref{eq.Etilphi} 
with the Fourier transform of \eqref{eq.Eh0} 
and write down the full solution in \eqref{clas_sol2A} in the form,
\[
\lim_{t\to + \infty}\, h_2( t, {\bf k}) \,-\, v \, =\, 
\frac{1}{\sqrt{2\omega_{\bf k}}}\left(f_{\bf k}\, e^{-i\omega_{\bf k} t}\,+\, 
\left\{ g_{-\bf k} \,+\,
2v \sqrt{2\omega_{\bf k}}\, e^{m\tau_\infty}\,(2\pi)^{3/2}\, \delta^{(3)}({\bf k})
\right\}e^{i\omega_{\bf k}t}\right).
\label{eq.Efull}
\]
Comparing with the the final-time boundary condition \eqref{eq.E3} we read off the expressions for the coefficient functions,
\begin{eqnarray}
 b_{\bf k}\,e^{\omega_{\bf k}T-\theta} &=& f_{\bf k} 
 \label{eq:Eb}
 \\
  b^*_{\bf k}&=& g_{-\bf k} \,+\, 2v \sqrt{2m}\, e^{m\tau_\infty}\,(2\pi)^{3/2}\, \delta^{(3)}({\bf k}) \,.
\label{eq:Ebd}
\end{eqnarray}
We will now make an educated guess that the parameter $T$ will be infinite in the limit $\varepsilon \to 0$.
In fact we will soon derive that $T= 3/(2 m\varepsilon)$, so this assumption will be justified \textit{a posteriori}. 
We can then re-write \eqref{eq:Eb} as
\[
 b_{\bf k}\,=\,  f_{\bf 0}\, e^{-\omega_{\bf k}T}\, e^{\theta}
\]
In the limit where $\varepsilon \to 0$, and thus $T\to\infty$, the factor $e^{-\omega_{\bf k}T}$ can be thought of as the 
regularisation of a momentum-space delta-function:
it cuts-off all non-vanishing values of ${\bf k}$ by minimising $\omega_{\bf k}$, thus reducing ${\bf k}$ to zero. 
Therefore, we set $f_{\bf k}$ to $ f_{\bf 0}$
in the equation above. 

Furthermore, since the function $b_{\bf k}$ is proportional to the (regularised) delta-function, its complex conjugate $ b^*_{\bf k}$
 must be too. This implies that the coefficient function $g_{-\bf k}$ in \eqref{eq:Ebd} 
must be zero \cite{Son:1995wz}, which verifies \eqref{eq:gk0}, so that \eqref{eq.Etilphi} becomes,
\[
\lim_{t\to + \infty}\, \tilde{\phi}( t, {\bf k})
  \,=\, 
\frac{1}{\sqrt{2\omega_{\bf k}}}  \,\, f_{\bf k}\, e^{-i\omega_{\bf k}t}\,.
\label{eq.Etilphi2}
\]

We have obtained the expression for the coefficient function $b_{\bf k}$ (and its complex conjugate) and 
also obtained a symbolic identity involving the parameters $T$, $\theta$  and the delta-function,
\[
b_{\bf k}\,=\, f_{\bf 0}\, e^{-\omega_{\bf k}T}\, e^{\theta} \,=\, 2v \sqrt{2m}\, e^{m\tau_\infty}\,(2\pi)^{3/2}\, \delta^{(3)}({\bf k}) \,=\, b^*_{\bf k}\,.
\label{eq:Esymb}
\]
This symbolic identity should be interpreted as follows. In the limit of strictly vanishing $\varepsilon$,
all these terms are proportional to the delta-function. Away from this limit, i.e. in the case of processes near
the multiparticle threshold where $0<\varepsilon\ll 1$, the function $\delta^{(3)}({\bf k})$ appearing in the third term above is not the strict delta-function, but a narrow peak with the singularity regulated by $\varepsilon$. This can be derived by
allowing the surface 
$\tau_\infty$ in the first term in \eqref{clas_sol2A} to be not completely flat at small non-vanishing $\varepsilon$, but to have a
tiny curvature $2\varepsilon/3\ll 1$ \cite{Son:1995wz}, thus leading to a regularised expression for $\delta^{(3)}({\bf k})$ in the
final term in \eqref{eq.Efull}.

To proceed, we integrate the two middle terms in \eqref{eq:Esymb} over $d^3k$,
\[
 f_{\bf 0}\, e^{\theta}\, \int d^3 k \, e^{-\omega_{\bf k}T} \,=\, 2v \, e^{m\tau_\infty}\,(2\pi)^{3/2}\,.
\label{eq:inteq}
\]
The integral on the left hand side of \eqref{eq:inteq},
\[
\int d^3 k \, e^{-\omega_{\bf k}T} \,=\, 4\pi\, m^3\, e^{-mT} \, \int_0^\infty dx\, x^2 \, e^{-mT(\sqrt{1+x^2}-1)}\,,
\]
where $x= k/m$ and note that this integral is dominated by $x\sim mT$, which at large $T$ allows us to 
simplify this as,
\[
4\pi\, m^3\, e^{-mT} \, \int_0^\infty dx\, x^2 \, e^{-mTx^2/2} \,=\, 4\pi\, m^3\, e^{-mT} \, \frac{\sqrt{\pi/2}}{(mT)^{3/2}}\,.
\]
We can now solve the equation~\eqref{eq:inteq} for $ f_{\bf 0}$ and 
 find that at large $T$,
\[
 f_{\bf 0}\,=\, \frac{4}{\sqrt{\lambda}}\, (T)^{3/2}\, e^{mT-\theta+m\tau_\infty}\,.
\label{eq:Esymb2}
\]

We can now compute the particle number $n$ and the energy $E$ in the final state using equations \eqref{eq:al4}
and the now known coefficient functions \eqref{eq:Esymb} along with \eqref{eq:Esymb2}. We find,
\begin{eqnarray}
n &=& 
\int d^3 k \,\, b_{\bf k}^* \, b_{\bf k}\, e^{\omega_{\bf k}T-\theta} 
\,=\,  \int d^3 k \,\, b_{\bf k}^* \, f_{\bf 0} 
\,=\, \frac{16}{\lambda} \, (2 \pi mT)^{3/2}\,e^{mT-\theta+2m\tau_\infty}
\label{eq:1stint}
\end{eqnarray}
and
\begin{eqnarray}
mn\varepsilon \,=\, E-mn &=&
 \int d^3 k \, \frac{{\bf k}^2}{2}\, b_{\bf k}^* \, b_{\bf k}\, e^{\omega_{\bf k}T-\theta} \nonumber\\ 
 &=&
 \int d^3 k \, \frac{{\bf k}^2}{2}\, b_{\bf k}^* \, f_{\bf 0}\,=\, 
\frac{16}{\lambda} \, (2 \pi mT)^{3/2}\,e^{mT-\theta+2m\tau_\infty}\, \frac{3}{2T}
\label{eq:2ndint}
\end{eqnarray}
It turned out that it was sufficient to know just the value of $f_{\bf k}$ at ${\bf k}=0$ to evaluate the integrals above,
due to the fact that $b_{\bf k}^*$ and $b_{\bf k}$ are sharply peaked at ${\bf k}=0$ as dictated by \eqref{eq:Esymb}.

Dividing the expression on the right hand side of \eqref{eq:2ndint} by the expression in \eqref{eq:1stint} we find,
\[
T\,=\, \frac{1}{m}\,\frac{3}{2}\, \frac{1}{\varepsilon}\,.
\label{eq:Tfound}
\]
The second parameter $\theta$ is found to be,
\[ 
\theta\,=\,
-\, \log\frac{\lambda n}{4}\,+\, \frac{3}{2}\log\frac{3\pi}{\varepsilon}\,+\, 2m\tau_\infty\,+\, \frac{3}{2}\,\frac{1}{\varepsilon}\,.
\label{eq:Thfound}
\]
We now finally substitute these parameters into the equation \eqref{eq:al52} for the `holy grail' function $W(E,n)$, and find, 
\begin{eqnarray}
W(E,n) &=&
ET \,-\, n\theta \,-\, 2{\rm Re} S_{E}[h]\,=\,mn(1+\varepsilon)T\,-\, n\theta \,-\, 2{\rm Re} S_{E}[h]
\nonumber\\\nonumber\\
&=& n\, \log\frac{\lambda n}{4}\,+\, n\left(\frac{3}{2} \log\frac{\varepsilon}{3\pi}+1\right)\,-\,  2nm\,\tau_\infty\,-\, 2{\rm Re} S_{E}[h]\,.
\label{eq:Eal5a}
\end{eqnarray}
We also note that the expression for $f_{\bf 0}$ found in \eqref{eq:Esymb2} evaluated with $T$ and $\theta$ given by 
\eqref{eq:Tfound}-\eqref{eq:Thfound}, reproduces the equation \eqref{eq:fk0}, which was our second constraint on the general
form of the solution 
$h_2(t_{\mathbb{C}},\bec{x})$ in \eqref{clas_sol2A}.

Before interpreting the expression \eqref{eq:Eal5a} for the `holy grail' function,
 we would like to separate the terms appearing on the right-hand side 
 into those that depend on the location and shape of the singularity surface $\tau_0(\bec{x})$, and those that do not.
The first two terms in \eqref{eq:Eal5a} have no dependence on the singularity surface; the third term, $2nm\,\tau_\infty$, depends
on its location at $\tau_\infty$.
The final term, $2{\rm Re} S_{E}$, is obtained by taking the real part of the three integrals appearing in \eqref{eq:Se}.
The first two integrals are along the Euclidean time $\tau$ segments of the contour and are real-valued,
\[
2{\rm Re} \, S^{(1,2)}_{E}\,=\,2\, \int  d^3 x \left[ -\, \int_{+\infty}^{\tau_0(\bec{x})} d\tau \, {\cal L}_{E}(h_1) \,-\,
\int_{\tau_0(\bec{x})}^0d\tau \, {\cal L}_{E}(h_2) \right]\,,
 \label{eq:Se12R}
\]
while the remaining integral along the third segment of the contour appears to be purely imaginary. This last statement is almost correct,
as it applies to the bulk contribution of the Minkowski-time integral $\int_{0}^\infty dt \, {\cal L}(h_2)$, but not to the boundary contribution 
at $t\to \infty$. The full contribution from the third segment of the contour is,\footnote{The expression \eqref{eq:Se3R} for the boundary contribution 
to the Minkowski action is also in agreement with the construction in \cite{Son:1995wz} and \cite{Libanov:1997nt}.}
\begin{eqnarray}
2{\rm Re} \, S^{(3)}_{E} &=& 2\, \int  d^3 x \left[ -\,i 
 \int_{0}^\infty dt \, \int d^3x\,  \partial_t\left(\tilde \phi \, \partial_t h_2\right) \right] \nonumber\\
 &=&
 -\,\int d^3 k \,\, b_{\bf k}^* \, b^{}_{\bf k}\, e^{\omega_{\bf k}T-\theta}\,\,=\, -\,n
 \,.
 \label{eq:Se3R}
\end{eqnarray}

Accounting for the effect of the boundary contribution \eqref{eq:Se3R}
we can write the expression for the rate \eqref{eq:Eal5a} in the form:
\[
W(E,n) \,=\, n \left(\log\frac{\lambda n}{4}\,+\, \frac{3}{2} \log\frac{\varepsilon}{3\pi}\,+\,\frac{1}{2}\right)
\,-\,  2nm\,\tau_\infty\,-\, 2{\rm Re} \,S^{(1,2)}_{E}(\tau_0)\,.
\label{eq:EalWfin1}
\]

This is a remarkable formula in the sense that the expression on the right-hand side of \eqref{eq:EalWfin1} cleanly separates into
two parts. The first part,
$n \left(\log\frac{\lambda n}{4}\,+\, \frac{3}{2} \log\frac{\varepsilon}{3\pi}\,+\,\frac{1}{2}\right)$, does not depend on the shape of the singularity surface $\tau_0(\bec{x})$ and coincides with the known
tree-level result for the scattering rate in the non-relativistic limit $0<\varepsilon\ll 1$, as we will demonstrate below. The 
entire dependence of $W(E,n) $ on $\tau_0(\bec{x})$ is contained in the last two terms in \eqref{eq:EalWfin1}, which correspond to the purely quantum
contribution in the $\varepsilon \to 0$ limit.

The tree-level contribution to $W$ is well-known; it was computed using the resummation of
Feynman diagrams by solving the tree-level recursion relations \cite{Libanov:1994ug} and integrating over the phase-space. In the model
\eqref{eq:L}, the tree-level result to the order $\varepsilon^1$ was derived in \cite{Khoze:2014kka} and reads,
\[
W(E,n; \lambda)^{\rm tree}  \,=\,n\,\left( f_1(\lambda n)\,+\, f_2(\varepsilon) \right)\,,
\label{eq:Wtree}
\]
where
\begin{eqnarray}
\label{f0SSB}
f_1(\lambda n)&=&  \log\left(\frac{\lambda n}{4}\right) -1\,, 
\\
\label{feSSB}
f_2(\varepsilon)|_{\varepsilon\to 0}&\to& f_2(\varepsilon)^{\rm asympt}\,=\, 
\frac{3}{2}\left(\log\left(\frac{\varepsilon}{3\pi}\right) +1\right) -\frac{25}{12}\,\varepsilon\,.
\end{eqnarray}
First ignoring the order-$\varepsilon^1$ terms in the tree-level contribution, we see that the perturbative result
is correctly reproduced by the 
first two terms in the semiclassical expression on the right-hand side of \eqref{eq:EalWfin1},
\[
W(E,n)^{\rm tree} \,=\, n \left(\log\frac{\lambda n}{4}-1\right)\,+\, \frac{3n}{2}\left( \log\frac{\varepsilon}{3\pi}+1\right)
\,.
\label{eq:EalWfinfin}
\]
Schematically, the contribution $n\log \lambda n \subset W^{\rm tree}$ comes from squaring the tree-level amplitude on threshold
and dividing by the Bose symmetry factor,
$\frac{1}{n!}\, (n! \lambda^{n/2})^2 \sim n!\lambda^n\sim e^{n\log \lambda n}$, while the contribution $\frac{3}{2}n\log \varepsilon$
comes from the non-relativistic $n$-particle phase space volume factor $\varepsilon^{\frac{3n}{2}} \sim e^{\frac{3}{2}n\log \varepsilon}$. 
[We refer the interested reader to Refs.~\cite{Libanov:1994ug,Khoze:2014kka} for more details on the derivation of $W(E,n)^{\rm tree}$
directly in perturbation theory.]

\medskip

The apparent agreement between the first term in the expression on the right-hand side of  \eqref{eq:EalWfin1}  and the result of an independent tree-level perturbative calculation \eqref{eq:EalWfinfin}, provides a non-trivial consistency check of the semiclassical formalism that led us to \eqref{eq:EalWfin1}.

Furthermore, it was shown in \cite{Son:1995wz} that the tree-level results are also correctly reproduced by the semiclassical result to order-$\varepsilon^1$.
It would also be interesting to pursue such 
terms at the quantum level, but this is beyond the scope of this paper.
We will neglect all ${\cal O}(\varepsilon)$ terms as they are vanishing in the $\varepsilon \to 0$ limit.

\medskip

We can finally re-write the expression \eqref{eq:EalWfin1} for the rate $W(E,n)$ in the form \cite{Son:1995wz},
\[
W(E,n) \,=\, W(E,n; \lambda)^{\rm tree}  \,+\, \Delta W(E,n; \lambda)^{\rm quant}\,,
\label{eq:EalWfin2}
\]
where the quantum contribution is given by
\begin{eqnarray}
\Delta W^{\rm quant} &=&
\,-\,  2nm\,\tau_\infty\,-\, 2{\rm Re} \,S^{(1,2)}_{E}
\nonumber\\ 
&=& 2nm\,|\tau_\infty|\,+\, 2 \int  d^3 x \bigg[ \int_{+\infty}^{\tau_0(\bec{x})} d\tau \, {\cal L}_{E}(h_1) \,+\,
\int_{\tau_0(\bec{x})}^0d\tau \, {\cal L}_{E}(h_2) \bigg]
\label{eq:EWq} \\
&=& 2nm\,|\tau_\infty|\,-\, 2 \int  d^3 x \bigg[ \int^{+\infty}_{\tau_0(\bec{x})} d\tau \, {\cal L}_{E}(h_1) \,-\,
\int_{\tau_0(\bec{x})}^0d\tau \, {\cal L}_{E}(h_2) \bigg] \nonumber.
\end{eqnarray}
Here we have used the fact that $\tau_{\infty}$ is manifestly negative (as the singularity surface away at $\bec{x} \neq 0$ is by construction assumed to be located at negative $\tau$) to indicate that $-2nm\,\tau_\infty$ is a positive-valued contribution $+2nm\,|\tau_\infty|$.

The problem of finding the singularity surface $\tau_0(\bec{x})$ that extremises the expression \eqref{eq:EWq} has a simple
physical interpretation \cite{Son:1995wz,Gorsky:1993ix,Khoze:2017ifq}: 
it is equivalent to finding the shape of the membrane $\tau_0(\bec{x})$
at equilibrium, which has the 
surface energy  ${\rm Re} \,S^{(1,2)}_{E}$ and is pulled at the point $\bec{x}=0$ by a constant force equal to $nm$.
Note that even before the extremisation of \eqref{eq:EWq}
with respect to $\tau_0(\bec{x})$, both configurations $h_1(x)$ and $h_2(x)$ are tightly constrained. They are required to be
solutions of the classical equations; they have to have satisfy the correct boundary conditions in time, and consequentially, their energy is fixed: $h_1$ has $E=0$ and $h_2$ has $E=nm$ (in the $\varepsilon \to 0$ limit). These conditions 
constrain the extremisation of \eqref{eq:EWq} with respect to $\tau_0(\bec{x})$.

\medskip
\section{Computing the rate: the thin-wall approximation}
\label{sec:thin_w}
\medskip

The main idea on which our calculation will be based is the geometrical interpretation of the saddle-point 
field configuration as a domain wall solution separating the vacua with  different VEVs $h \to \pm v$ on the different
sides of the wall. Our scalar theory with the spontaneous symmetry breaking in \eqref{eq:L} clearly supports
such field configurations.
The solution is singular on the surface of the wall, and the wall thickness is $\sim 1/m$. The effect of the `force'
$nm$ applied to the domain wall locally pulls upwards  the centre of the wall and gives it a profile $\tau_0({\bf x})$
depicted in Fig.~\ref{fig:tau}.
When computing the Euclidean action on the solution characterised by the
domain wall at $\tau_0({\bf x})$, it will be represented by the action of a thin-wall bubble. The shape of the bubble will
be straightforward to determine by extremizing the action in the thin-wall approximation, and the validity of this approximation 
will be be justified in the limit $\lambda n \to \infty$.\footnote{The idea to use of the thin-wall approximation in the large $\lambda n$ limit
was pursued earlier by Gorsky and Voloshin in Ref.~\cite{Gorsky:1993ix} where it was applied to the standard regular bubbles of the
false vacuum that were interpreted as intermediate physical bubble states in the process $1^* \to {\rm Bubble} \to n$. Conceptually, 
this is different from
our approach where the thin-wall solutions are singular points on the deformed contours of the path integral; they cannot be 
obviously interpreted 
as physical  macroscopic states supposedly occurring as intermediate states in the $1^* \to n$ process.}

Our first task is to implement the realisation of the singular field configuration $h(x)$ in terms of domain walls with thin-wall singular surfaces.
The $h_1$ branch of the solution is defined on the first part of the time-evolution contour, i.e. the imaginary time interval
$+\infty > \tau \ge \tau_0(\bec{x})$.
It is given by,
\[
h_1(\tau,\bec{x}) \,=\, h_{0E} (\tau-\tau_0(\bec{x})) \,+\, \delta h_1(\tau,\bec{x})\,.
\label{clas_sol111Aag}
\]
The first term on the right-hand side of \eqref{clas_sol111Aag} is the familiar singular domain wall,
\[
h_{0E} (\tau-\tau_0(\bec{x})) \,=\, v \left(\frac{ 1\,+\,e^{-m (\tau-\tau_0(\bec{x}))}}
{1\,-\, e^{-m (\tau-\tau_0(\bec{x}))}}\right)\,,
\label{clas_sol222inf}
\]
with its centre (or position) at $\tau = \tau_0(\bec{x})$.
This profile is similar to the one depicted in Fig.~\ref{fig:kink}, the field configuration interpolates between $h=+v$ at $\tau \gg  \tau_0(\bec{x}) $ and  $h=-v$ at $\tau \ll  \tau_0(\bec{x})$,
and is singular on the 3-dimensional surface $\tau=\tau_0(\bec{x})$. Since $\tau_0(\bec{x})$ depends on the spatial variable,
the correction $\delta h_1(\tau,\bec{x})$ is required in \eqref{clas_sol111Aag} to ensure that the entire field configuration $h_1(x)$ 
satisfies the classical equations. The $\delta h_1$ term vanishes on the singularity surface; in fact it is straightforward to show that 
$\delta h_1 \sim (\tau-\tau_0(\bec{x}))^3$ near the singularity surface by solving the linearised classical equations for $\delta h_1$ 
in the background of the singular $h_0$ \cite{Son:1995wz}.
The initial time condition on $h_1$ is
\[
\lim_{\tau\to \infty} h_1 (x)\,=\, v + {\cal O}(e^{-m\tau})\,,
\]
which also guarantees that $\delta h_1(x)\to 0$ exponentially fast at large $\tau$.
Hence, in computing the action integral of $h_1(x)$ in the thin-wall approximation, where the main contribution comes from $\tau$ in the vicinity
of $\tau_0(\bec{x})$, it will be a good approximation to neglect $\delta h_1(x)$ and 
use,
\[
{\rm thin\,  wall}\, : \quad h_1(\tau,\bec{x}) \,\approx\, h_{0E} (\tau-\tau_0(\bec{x})) \,.
\label{eq:thwapp1}
\]

Now consider the second branch of the solution, $h_2(x)$. We search for solutions of the form required by Eq.~\eqref{clas_sol2A},
\[
h_2(t_{\mathbb{C}},\bec{x}) \,=\, h_0 (t_{\mathbb{C}}) \,+\, \tilde{\phi}(t_{\mathbb{C}},\bec{x})
\,,
\label{clas_sol2Aag}
\]
The first term on the right-hand side of \eqref{clas_sol2Aag} is the uniform in space and singular on the plane $\tau=\tau_{\infty}$
classical configuration
\[
 h_0 (t_{\mathbb{C}}) \,=\, v\, \left(\frac{ 1\,+\,e^{im (t_{\mathbb{C}}-i\tau_\infty)}}
{1\,-\, e^{im (t_{\mathbb{C}}-i\tau_\infty)}}\right)
\,.
\label{clas_sol22inf}
\]
In the previous section we derived the asymptotic form for the second term, $\tilde{\phi}(t_{\mathbb{C}},\bec{x})$,  
appearing on the right-hand side of \eqref{clas_sol2Aag}: for the final part of the time-evolution contour, where 
$t_{\mathbb{C}} = t \to +\infty$ we have,
\[
\lim_{t\to + \infty}\, \tilde{\phi}( t, {\bf x})
  \,=\, \int\frac{ d^3 k}{(2\pi)^{3/2}}\frac{1}{\sqrt{2\omega_{\bf k}}}  \,\, f_{\bf k}\, e^{-i\omega_{\bf k}t} \,,
  \label{eq:tilphiinf}
\]
This is in agreement with
Eqs.~\eqref{eq.Eh0} and \eqref{eq.Etilphi2} and its characteristic feature is that it contains only the negative frequency components
(at large $t$).
The coefficients of positive frequency components that were present in $\tilde{\phi}( t, {\bf x})$ at earlier times, closer to the origin at 
$t\sim 0$ become 
suppressed as the real time variable $t$ grows and ultimately disappear for a sufficiently large positive $t$.
We are now going to assume that the asymptotic expression \eqref{eq:tilphiinf} which is valid in the $mt \gg 1$ regime on or near
the real time axis, in fact also continues to hold when $\tilde{\phi}( t+i\tau, {\bf x})$ moves in the $\tau$ direction, i.e. perpendicular
to the real time contour at large fixed value of $t$.
More precisely we expect that the equation \eqref{eq:tilphiinf} generalises to the complex time variable $t_{\mathbb{C}}$ 
and holds as long as the real time coordinate $t$ is large ($t\gg 1/m$),
\[
\lim_{ t \to + \infty}\, \tilde{\phi}( t_{\mathbb{C}}, {\bf k})\,=\,
 \frac{1}{\sqrt{2\omega_{\bf k}}}  \,\,  f_{\bf k}\, e^{-i\omega_{\bf k}t_{\mathbb{C}}} \,=\,
 \frac{1}{\sqrt{2\omega_{\bf k}}}  \,\,  f_{\bf k}\, e^{\omega_{\bf k}\tau} \, e^{-i\omega_{\bf k}t} 
   \,.
  \label{eq:tilphiinf2}
\]
As always,  $t_{\mathbb{C}}=t+i\tau$, and for concreteness we will take 
the $\tau$ component to be negative, i.e. we will only need this expression for shifting downwards from the real time 
contour at large $t$.

\medskip

 \begin{figure*}[t]
\begin{center}
\includegraphics[width=0.95\textwidth]{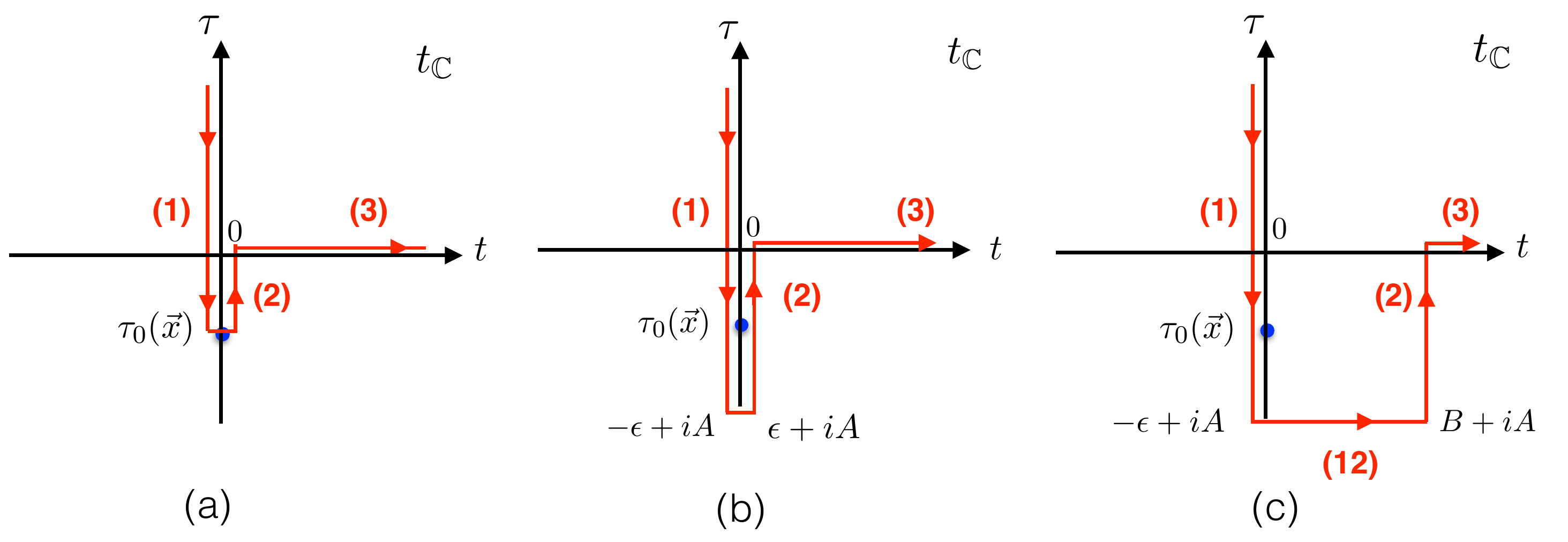}
\end{center}
\vskip-.5cm
\caption{Deformations of the time evolution contour in $t_{\mathbb{C}}$.  {\bf Plot (a)} shows the
original contour that touches the sigularity located at $t=0$, $\tau=\tau_0(\bec{x})$.
{\bf Plot (b)} gives the resolved contour, now surrounding the singularity with the vertical segments of the contour shifted infinitesimally by
 $\pm i \epsilon$ and descending to $\tau = -A$.
{\bf Plot (c)} shows now a finite deformation of the vertical part (2) of the contour the right. We use large shift values, $-A\gg 1/m$ and $B\gg 1/m$ to justify the thin wall approximation. Consequitive contour segments are denoted (1), (12), (2) and (3).}
\label{fig:contour3}
\end{figure*}

We now turn to the evaluation of the Euclidean action integrals appearing in \eqref{eq:Se12R} and \eqref{eq:EWq}. 
On the first segment
of the contour, indicated as (1)  in Fig.~\ref{fig:contour3}~(a), the classical field configuration is $h_1(x)$, while on the
segment (2) of the contour in Fig.~\ref{fig:contour3}~(a), the field is $h_2(x)$, hence,
\[
{\rm Fig.~\ref{fig:contour3} (a)}: \quad
-\,{\rm Re} \, S^{(1,2)}_{E}\,=\, \int  d^3 x \left[ \int_{+\infty}^{\tau_0(\bec{x})} d\tau \, {\cal L}_{E}(h_1) \,+\,
\int_{\tau_0(\bec{x})}^0d\tau \, {\cal L}_{E}(h_2) \right].
 \label{eq:Se12R2}
\]
The two individual integrals in \eqref{eq:Se12R2} are singular at the integration limit $\tau=\tau_0(\bec{x})$. 
However, their sum is expected to be finite, which is also known from in the Landau-WKB approach in
 Quantum Mechanics \cite{Landau}. 

Instead of reaching the singularity and then cancelling 
the resulting infinite contributions 
at $\tau\to \tau_0(\bec{x})$, we advocate a more practical approach and deform the integration contour to encircle the singularity,
as shown in the contour deformation from Fig.~\ref{fig:contour3} (a) to  Fig.~\ref{fig:contour3} (b).
The contour is shifted infinitesimally by $t=-\epsilon$
in the first integral in \eqref{eq:Se12R2} and by $t=+\epsilon$ in the second. 
Since the integration contour in Fig.~\ref{fig:contour3} (b) passes on either side of the singularity at $\tau=\tau_0(\bec{x})$, the action
integrals and the solutions themselves are finite. One can extend the integration contours down to $\tau=-\infty$ or to 
any arbitrary value $\tau=-A$. 
At $\tau=-A$, where $\tau$ is well below the final singularity surface $\tau_\infty$, the two contours are joined. As a result,  
the action integrals now read:
\[
{\rm Fig.~\ref{fig:contour3} (b)}: \quad
-\,{\rm Re} \, S^{(1,2)}_{E}\,=\,
\int_{\,+\infty-i\epsilon}^{\,-A-i\epsilon} d\tau \, L_{E}[h_1] \,+\,
\int_{\,-A+i\epsilon}^{\,0+i\epsilon}d\tau \, L_{E}[h_2] \,,
 \label{eq:Se12R22}
\]
where $L_{E}=\int d^3x\, {\cal L}_{E}$, and each of the two integrals in \eqref{eq:Se12R22} is finite.
The first integral in \eqref{eq:Se12R22} depends on the classical branch $h_1(x)$, and in the thin wall
approximation \eqref{eq:thwapp1}
we will be able to evaluate it as the functional of the surface $\tau_0(\bec{x})$ using the $h_{0E}$ profile in \eqref{clas_sol222inf}.

The second integral in \eqref{eq:Se12R22} is evaluated on the classical configuration $h_2(x)$. It is given by 
\eqref{clas_sol2Aag}, where the correction $\tilde{\phi}( t_{\mathbb{C}}, {\bf x})$ to the classical profile 
$h_0 (t_{\mathbb{C}}) $ in \eqref{clas_sol22inf} is known at large values of the
$t$, see Eq.~\eqref{eq:tilphiinf2}. To make use of these expressions for $h_2(x)$ we continue 
shifting the contour to the right by a 
constant value $B$ as shown in Fig.~\ref{fig:contour3} (c).
The resulting contributions to the Euclidean action from the integration contour in Fig.~\ref{fig:contour3} (c)
are given by the following integrals,
\[
{\rm Fig.~\ref{fig:contour3} (c)}: \quad
-\,{\rm Re} \, S^{(1,12,2)}_{E}\,=\,
\int_{\,+\infty-i\epsilon}^{\,-A-i\epsilon} d\tau \, L_{E}[h_1] \,+\,
i\int_{(12)}d t  \, L[h_2] \,+\,
\int_{\,-A+iB}^{\,0+iB}d\tau \, L_{E}[h_2] \,.
 \label{eq:Se12R22c}
\]
An obvious consequence of the thin wall approximation is that the middle integral on the right hand side
of \eqref{eq:Se12R22c} vanishes for $A$ sufficiently far below $\tau_\infty$ since in this case we are sufficiently deep
into the $h_2 = -v$ domain, the field configuration is constant there and the action on the (12) segment of the contour vanishes,
$\int_{(12)}d t  \, L[h_2]\,=\, 0$.

 \begin{figure*}[t]
\begin{center}
\includegraphics[width=0.6\textwidth]{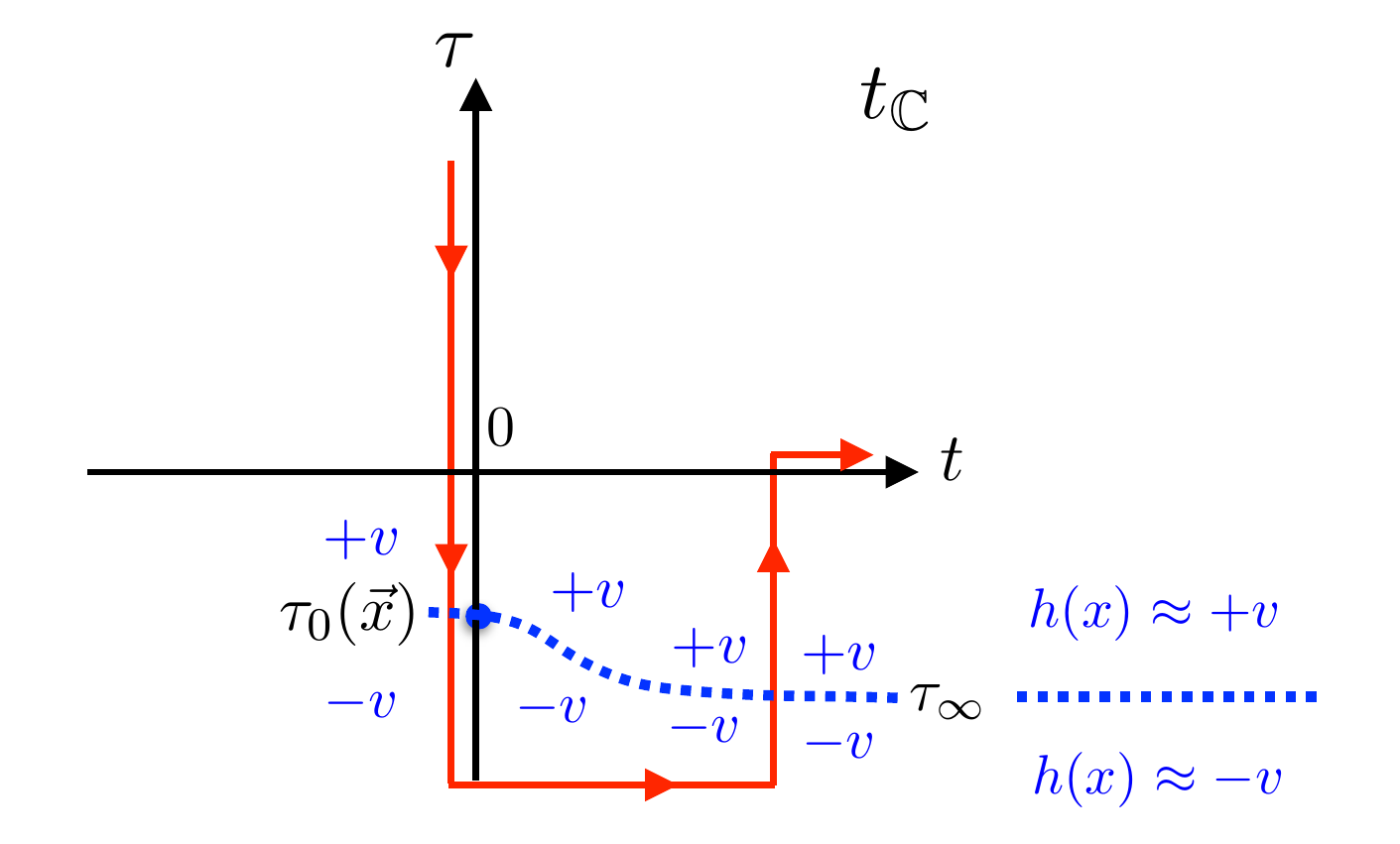}
\end{center}
\vskip-.5cm
\caption{The same complex-time evolution contour as in Fig.~\ref{fig:contour3} (c). The boundary separating 
the domains $h(x) \to +v$ and 
$h(x)\to -v$ for the classical solution in a thin wall approximation is shown as the dotted blue line.The singularity of the solution is 
at the point $t=0$, $\tau=\tau_0({\bf x})$ and depicted as a blob on the dotted line of the inter-domain boundary.}
\label{fig:contour_fin}
\end{figure*}

\medskip

Next, we can readily evaluate the last integral in \eqref{eq:Se12R22c}. It arises from segment (2) of the contour
in Fig.~\ref{fig:contour3} (c), which is the integral over the imaginary time component $d\tau$ and is
situated at a fixed value of real time at ${\rm Re} \, t_{\mathbb{C}} =B \gg 1/m$. 
Hence we can use the 
asymptotic expression~\eqref{eq:tilphiinf2} for  $\tilde{\phi}( t_{\mathbb{C}}, {\bf x})$ on this segment of the contour,
so that the entire solution $h_2(x)$ is given by,
where
\[
h_2^{\rm segment\, (2)}\,=\,   v\, \left(\frac{ e^{-im B -m(|\tau|-|\tau_\infty|)}\,+\,1}
{e^{-im B -m(|\tau|-|\tau_\infty|)}\,-\,1}\right)
\,+\,  \int\frac{ d^3 k}{(2\pi)^{3/2}}\,
 \frac{1}{\sqrt{2\omega_{\bf k}}}  \,\, f_{\bf k}\, e^{-\omega_{\bf k}|\tau|} \, e^{-i\omega_{\bf k}B} 
   \,.
  \label{eq:tilphiinf22}
\]
Note that on this segment of the contour $t=B$, $-A\le \tau \le 0$, hence 
$0\le |\tau|\le A$ and $0<|\tau_\infty|\ll A$.
In the large $\lambda n$ limit, we will find in the following section that in fact  $0\ll |\tau_\infty|$, 
and we find that the 
only non-trivial contribution in the thin wall limit on this segment of the contour will come from
the first term on the right hand side of \eqref{eq:tilphiinf22}. The location of the wall
 separating the two $\pm v$ domains of the field configuration is depicted in 
 Fig.~\ref{fig:contour_fin}.
The $\tilde{\phi}$ term on its own cannot
contribute to the action integral since it contains only the negative frequencies, at the same time,
its overlap with the $h_0$ configuration 
 at $\tau \approx \tau_\infty$ is exponentially suppressed by $e^{-m|\tau|_\infty} \ll 1$.  Hence only we have, 
\[
{\rm thin\,  wall}\, : \quad h_2(\tau,\bec{x}) \,\approx\, h_{0E} (\tau-\tau_\infty) \,.
\label{eq:thwapp2}
\]
This equation is applicable on the segment (2) of the contour in Fig.~\ref{fig:contour3}~(c), and 
 the argument $\tau$ of the both functions in \eqref{eq:thwapp2} is understood as $\tau - iB$.
 
 Equations \eqref{eq:thwapp1} and \eqref{eq:thwapp2} give us the required precise implementation of the thin wall
 approximation that we will apply in what follows.
 In both cases the field configurations, $h_1$ in \eqref{eq:thwapp1}, and $h_2$ in \eqref{eq:thwapp1}, are approximated
 in the thin wall approach
 by the Brown's solution profile $h_{0E}$. The important difference between the two cases, however, is that the  
 domain wall in \eqref{eq:thwapp1} is the ${\bf x}$-dependent surface $\tau_0(({\bf x})$, while in the case of the
 $h_2$ configuration in \eqref{eq:thwapp2}, the domain wall is at $\tau_\infty$ and is spatially-independent.
 As the result, the the first integral on the right hand side of our expression for the action in \eqref{eq:Se12R22c},
 is the functional of the domain-wall surface $\tau_0(({\bf x})$,
 \[
 S^{(1)}_{E}\,=\,
\int^{\,+\infty-i\epsilon}_{\,-A-i\epsilon} d\tau \, L_{E}[h_1]\,=\, S_{E}[\tau_0(({\bf x})]\,,
 \label{eq:Se12R22c1}
\]
while the the third integral in \eqref{eq:Se12R22c} is evaluated on the uniform in space solution \eqref{eq:thwapp2}
and is a constant,
 \[
 S^{(2)}_{E}\,=\,
-\, \int_{\,-A+iB}^{\,0+iB}d\tau \, L_{E}[h_2] \,=\, -\,{\rm const}\,.
 \label{eq:Se12R22c2}
\]
In both cases, on the segment (1) and the segment (2) of the contour, the field configurations are regular,
as, by construction, the contour avoids the singularity by the $-i \epsilon$ shift in the first integral and by the
$+iB$ shift in the second.
 
We now proceed to compute the integral in \eqref{eq:Se12R22c2}.
This integral is evaluated on the field configuration,
\[
h_{2} (\tau+iB) \,=\, v \left(\frac{ 1\,+\,e^{-m (\tau-\tau_\infty+iB)}}
{1\,-\, e^{-m (\tau-\tau_\infty+iB)}}\right)\,,
\label{clas_sol22infh1}
\]
and can be calculated exactly\footnote{For simplicity we extend the integration limits 
along the vertical axis to $\pm \infty$. Given the narrow width of the wall, any changes due to this extension are negligible.}, 
giving,
\[
\int^{+\infty+iB}_{-\infty+iB} d\tau \int  d^3 x {\cal L}_{E}(h_2) \,=\,
 \mu \int_0^R 4\pi r^2 dr\,=\, 
\mu\,  \frac{4\pi}{3} \, R^3\,.
 \label{eq:Se1act}
\]
Since the field is uniform in space, to ensure that the $\int d^3x$ is finite,
 we used the finite volume regularisation with finite spatial radius $R$.     
 The infinite-volume limit, $R\to \infty$, will be taken at the end of the calculation, 
after combining the two action integrals in \eqref{eq:Se12R22c1} and \eqref{eq:Se1act}.
The parameter $\mu$ appearing on the right-hand side of \eqref{eq:Se1act}
is the surface tension on the bubble solution \eqref{clas_sol22infh1},
\[
\mu\,=\, \int_{-\infty+i\epsilon}^{+\infty+i\epsilon} d\tau \left( \frac{1}{2} \left(\frac{d h}{d\tau}\right)^2 +
 \frac{\lambda}{4} \left( h^2 - v^2\right)^2 \right) \,=\,  \frac{m^3}{3 \lambda}\,.
 \label{eq:mu}
\]
It can easily be checked (e.g. by use of the residue theorem)
that the value of $\mu$ does not depend on the numerical value of $iB$ in the shift of the integration contour:
any value of $iB\neq 0$ that shifts the contour such that it does not pass directly through the singularity at $\tau_\infty$  is fine.
This shift-independence argument also applies to the integral on the fist segment of the contour where the shift is $-i\epsilon$.

\medskip

Let us summarise our construction up to this point. We have derived the expression for the contribution of quantum effects \eqref{eq:EWq}
to the semiclassical
rate $W$ \eqref{eq:EalWfin2} in the form,
\[
\frac{1}{2} \Delta W^{\rm quant}\,=\, 
nm\,|\tau_\infty|\,-\,  
\underbrace{\int^{\,+\infty+i\epsilon}_{\,-\infty-i\epsilon} d\tau \, L_{E}(h_1;\tau_0(\bec{x}))}_{\equiv \, S_{E}[\tau_0(\bec{x})]} 
\,+\,\, \frac{4\pi}{3} \,\mu R^3
\,.\label{eq:halfEWqnew}
\]
We note that no extremisation of the rate with respect to the surface $\tau=\tau_0(\bec{x})$ has been carried out so far.
The expression in \eqref{eq:halfEWqnew} is the general formula equivalent to the expression in \eqref{eq:EWq}.
It will be now extremised with respect to the domain wall surface $\tau_0(\bec{x})$. 
The constant term $\frac{4\pi}{3} \,\mu R^3$ will be cancelled with its counterpart arising from the action integral in \eqref{eq:halfEWqnew} 
before the infinite-volume limit is taken.

\medskip

Following from the discussion at the end of section~\ref{sec:son_loops}, the shape of the singular surface, $\tau_0(\bec{x})$, should be determined by
extremising the function $\Delta W^{\rm quant}$ in the exponent of the multiparticle probability rate.
This is equivalent to
searching for a stationary (i.e. equilibrium surface) configuration described by the `surface energy' functional, given by the right hand side of  \eqref{eq:halfEWqnew}.
Finding the stationary point corresponds to balancing the surface energy of the stretched surface, given by the integral 
$S_{E}[\tau_0(\bec{x})]$ in \eqref{eq:halfEWqnew},
against the force $nm$ that stretches the surface $\tau_0(\bec{x})$ by the
amount $|\tau_\infty|$. The third term on the right hand side of \eqref{eq:halfEWqnew} plays no role in the extremisation procedure over $\tau_0(\bec{x})$ and gives a positive-valued constant 
contribution to $\frac{1}{2} \Delta W^{\rm quant}$ that will be cancelled against its counterpart in $S_{E}[\tau_0(\bec{x})]$.
The overall result will be finite, as expected in the infinite volume limit.

\medskip

The action $S_{\rm Eucl}[\tau_0({\bf x}]$ can now be written as an integral over the domain wall surface $\tau_0({\bf x})$
in the thin-wall approximation. This is equivalent to stating that
 the action is equal to the surface tension of the domain wall $\mu$ already computed in \eqref{eq:mu}
 times the area. The infinitesimal element
 of the 3-dimensional area of a surface curved in 3+1 dimensions 
 is $4\pi \mu \, r^2 \sqrt{(d\tau)^2 +(dr)^2}$. Hence the action reads,
 \[ 
 S_{\rm Eucl}[ \tau_0(r)] \,=\,
\int_{\tau_\infty}^0 d\tau \,4\pi \mu \, r^2 \sqrt{1+\dot r^2}\,\,\equiv\,  \int_{\tau_\infty}^0 d\tau \, L (r,\dot r)
\,,
\label{eq_thinw}
\]
where $r=|{\bf x}|$ and  $\dot r= dr/d \tau$. The integral depends on the choice of the domain wall surface 
$\tau_0({\bf x})$
implicitly via dependence on $\tau$ of $r(\tau)$ and $\dot r(\tau)$ which are computed on the domain wall.
 
Since  $L (r,\dot r)$ has the meaning of the Lagrangian, we can introduce the  Hamiltonian function 
defined in the standard way\footnote{In Euclidean space
$L=K+P$ and $H=P-K$ where $K$  and $P$ are the kinetic and potential energies respectively.} as the Legendre transformation,
\[
H(p,r)\,=\, L (r,\dot r) \,-\, p\, \dot r\,,
\label{Hdef}
\]
where the momentum $p$, conjugate to the coordinate $r$, is 
\[
p\,=\, \frac{\partial L (r,\dot r) }{\partial \dot r}\,\,\,=\, 4\pi\, \mu \frac{r^2 \dot r}{\sqrt{1+\dot r^2}}
\label{eq:defp1}
\]
On a classical trajectory $r=r(\tau)$ that satisfies the Euler-Lagrange equations corresponding to $L (r,\dot r)$,
the Hamiltonian is time-independent, $dH/d \tau=0$, and is given by the energy $E$ of the classical trajectory $r=r(\tau)$.\footnote{It is important
not to confuse the energy of the classical trajectory $r=r(\tau)$ -- which is essentially the Euclidean surface energy of the domain wall --
with the energy of the classical solutions $h_1$ and $h_2$. Both energy variables are denoted as $E$, but the energy of the domain wall
at the stationary point will turn out to be $E=mn$ while the energy of the corresponding field configuration $h_1$ was $E=0$.}
Hence, on a stationary point of  $S_{\rm Eucl}[ \tau_0(r)]$ that has the energy $E$ we can rewrite the action as
\[ 
S_{\rm Eucl}[ \tau_0(r)]_{\rm stationary} \,=\, -\tau_{\infty}\,E
\,+\,
 \int_{\tau_\infty}^0 d\tau \, (L - H) \, 
\,=\,  -\,E\tau_{\infty} \,+\, \int_{R}^0 p(E) \,dr
\,.
\]
Here we added and subtracted the constant energy of the solution $E=H$ in the integral, used the fact that $L-H=p \dot{r}$ 
and have set the lower and upper integration limits at $r(\tau_\infty)=R$ and $r(0)=0$. 
The expression above gives us $S_{\rm Eucl}[ \tau_0(r)]$ on a trajectory $r(\tau)$, or equivalently $\tau=\tau_0(r)$ 
which is a classical trajectory i.e. an extremum of the action for a fixed energy $E$.
Equivalently, for the stationary point of the expression in \eqref{eq:halfEWqnew} we have,
\[
\frac{1}{2} \Delta W^{\rm quant} \,=\, 
(E-nm)\tau_\infty\,-\,   \int_{R}^0 p(E) \,dr
\,+\, \frac{4\pi}{3} \,\mu R^3
\,.\label{eq:halfEWqst}
\]
Extremization of this expression with respect to the parameter $\tau_\infty$ gives $E=nm$ thus selecting the
energy of the classical trajectory to be set at $nm$ as required,
\[
\frac{1}{2} \Delta W^{\rm quant}_{\quad\rm stationary} \,=\, 
  -\, \int_{R}^0 p(E) \,dr
\,+\, \frac{4\pi}{3} \,\mu R^3 \,, \qquad E=nm 
\,.\label{eq:halfEWqfi}
\]

To evaluate \eqref{eq:halfEWqfi} we need to determine the dependence of the momentum of the classical trajectory on its energy.
To find $p(E)$, we start by writing the expression for the energy, $E= L-p\dot{r}$, in the form
\[
E\,=\, 4\pi \mu \, r^2 \sqrt{1+\dot r^2} \,-\, 4\pi\, \mu \frac{r^2 \dot r}{\sqrt{1+\dot r^2}}\,=\,
4\pi\, \mu \frac{r^2}{\sqrt{1+\dot r^2}}\,,
\label{eq:Econs}
\]
and then compute the combination $E^2 + p^2$ using the above expression and \eqref{eq:defp1},
\[
E^2\,+\, p^2\,=\, \left(4\pi \mu \,r^2 \right)^2 \left( \frac{1}{1+\dot r^2}\,+\, \frac{\dot r^2}{1+\dot r^2}\right)\,=\,
 \left(4\pi \mu \,r^2 \right)^2\,.
 \]
This gives the desired expression for the momentum $p=p(E)$,
\[
p(E,r) \,=\,-\,  4\pi \, \mu \, \sqrt{r^4- \left(\frac{E}{4\pi\mu}\right)^2}\,,
\label{eq:conjp}
\] 
where have selected in \eqref{eq:conjp} the negative root for the momentum
in accordance with the fact that $p(\tau) \propto \dot{r}$ (as follows from \eqref{eq:defp1})
and that $r(\tau)$ is a monotonically decreasing function.

Substituting this into the expression \eqref{eq:halfEWqfi} we have,
\[
\frac{1}{2} \Delta W^{\rm quant} \,=\, 
-  \int_{R}^{r_0} p(E) \,dr
\,+\, \frac{4\pi}{3} \,\mu R^3\,=\, 
-\, \int^{R}_{r_0}  
4\pi \, \mu \, \sqrt{r^4-r_0^4} \,dr
\,+\, \frac{4\pi}{3} \,\mu R^3
\,.\label{eq:pdrN}
\]
The minimal value of the momentum (and the lower bound of the integral in \eqref{eq:pdrN}) is cut-off at the critical radius $r_0$,
\[
r_0^2 \,=\, \frac{E}{4\pi\mu} \,,
\label{r0def}
\]
Below we will also consider the contribution to the integral \eqref{eq:pdrN} on the interval $0\le r \le r_0$ but for now we will temporarily ignore it.

The integral on the right hand side of \eqref{eq:pdrN} is evaluated as follows,
\[
\int_1^{R/r_0}
 \sqrt{x^4-1} \,\,dx\,=\, \left[\frac{1}{3}\, x\, \sqrt{x^4-1}\,- \, \frac{2}{3} \, i\,  {\rm EllipticF}[{\rm ArcSin}(x), -1]\right]^{x=R/r_0}_{x=1}
 \nonumber
 \]
 where the {\it Mathematica} function ${\rm EllipticF}[z,m]$ is also known as the elliptic integral of the first kind $F(z|m)$.
The integral simplified in the $R/r_0 \to \infty$ limit giving,
\begin{eqnarray}
(-4\pi \mu r_0^3)\, \int_1^{R/r_0}  \sqrt{x^4-1} \,\,dx &\to& 
-\, \frac{4\pi}{3} \,\mu R^3
\,+\, 4\pi \mu r_0^3\, \sqrt{4\pi}\,
\frac{1}{3}\, \frac{\Gamma(5/4)}{\Gamma(3/4)}
\nonumber\\
&=& -\, \frac{4\pi}{3} \,\mu R^3
\,+\, \frac{E^{3/2}}{\sqrt{\mu}}\, 
\frac{1}{3}\, \frac{\Gamma(5/4)}{\Gamma(3/4)}\,.
\label{eq:pdrNev}
\end{eqnarray}
We see that the large volume constant term $\frac{4\pi}{3} \,\mu R^3$ cancels between the expressions in \eqref{eq:pdrNev}
and \eqref{eq:pdrN}, as expected. The final result for the thin-wall trajectory contribution to the quantum rate is given by,
\[
 \Delta W^{\rm quant} \,=\, 
\frac{E^{3/2}}{\sqrt{\mu}}\, 
\frac{2}{3}\, \frac{\Gamma(5/4)}{\Gamma(3/4)} \,=\, 
\, \frac{1}{\lambda} \, (\lambda n)^{3/2}\, \frac{2}{\sqrt{3}}\,
\frac{\Gamma(5/4)}{\Gamma(3/4)}\,\simeq\, \, 0.854\,  n \sqrt{\lambda n}
\,.\label{eq:pdrNfinalr0}
\]
We note that this expression is positive-valued, that it grows in the limit of $\lambda n \to \infty$, and that it has
the correct scaling properties for the semiclassical result, i.e. it is of the form $1/\lambda$ times a function of $\lambda n$.

Our result \eqref{eq:pdrNfinalr0} reproduces the expression derived in our earlier paper \cite{Khoze:2017ifq}
and is also in agreement with the expression derived even earlier in Ref.~\cite{Gorsky:1993ix}.

It also follows that the thin-wall approximation is fully justified in the $\lambda n \gg 1$ limit
as originally noted in ~\cite{Gorsky:1993ix,Khoze:2017ifq}.
The thin-wall regime corresponds to the radius of the bubble being much greater than the thickness of the wall,
$r \gg 1/m$. In our case the radius is always greater than the critical radius,
\[
r m\, \ge\, r_0 m\,=\, m\left( \frac{E}{4\pi \mu}\right)^{1/2} \propto\, \left( \frac{\lambda\, E}{m}\right)^{1/2} =\,
 \sqrt{\lambda n}\,\,\gg\, 1\,,
\]
where we have used the value for the energy  $E=nm$ on our solution.

\medskip

\begin{figure*}[t]
\begin{center}
\includegraphics[width=0.6\textwidth]{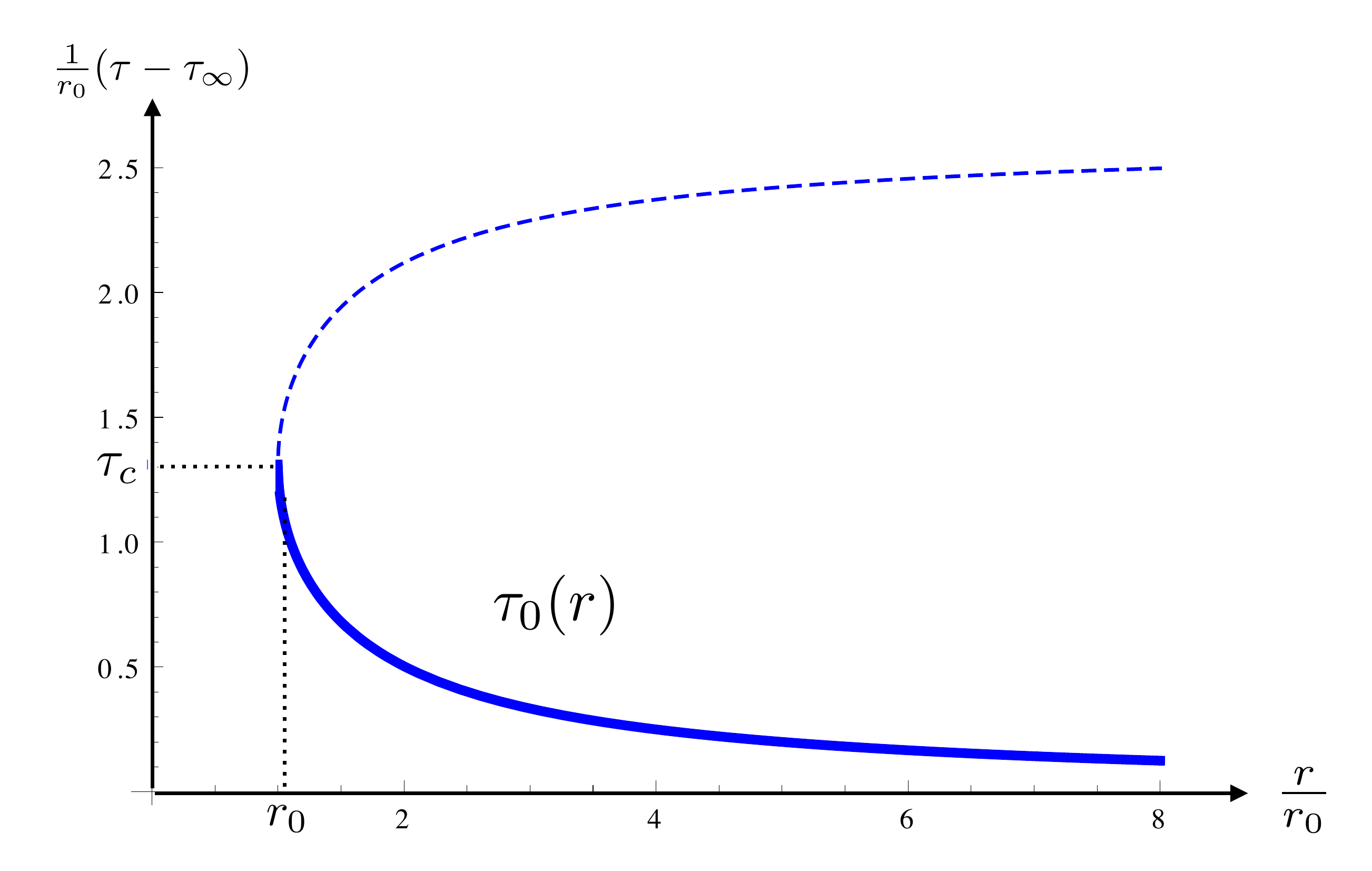}
\end{center}
\vskip-.5cm
\caption{Extremal surface $\tau=\tau_0(r)$ of the thin wall bubble solution \eqref{eq:profile1}. Solid line denotes the 
bubble wall profile of the bubble radius $r$ above the critical radius $r_0$. The dashed line corresponds to the branch
of the classical trajectory beyond the turning point at $r_0$. }
\label{fig:profile}
\end{figure*}

One can ask what is the actual classical trajectory $r(\tau)$ or equivalently the wall profile $\tau=\tau_0(r)$ of the classical bubble
on which the rate $W$ was computed in \eqref{eq:pdrNfinalr0}. To find it we can integrate the equation for the conserved energy
\eqref{eq:Econs}
on our classical solution,
\[
E\,=\, 4\pi\, \mu \frac{r^2}{\sqrt{1+\dot r^2}}\,,
\] 
or, equivalently, the expression $(r/r_0)^4 \,=\, 1+ \dot{r}^2$. One finds,
\[
\int_{\tau_{\infty}}^{\, \tau} d\tau\,=\,-\, \int_R^{\,r} \frac{dr}{\sqrt{\left(\frac{r}{r_0}\right)^4-1}}\,,
\]
which after integration can be expressed in the form,
\[
\tau(r) \,=\, \tau_\infty  \,+\, r_0 \left(\frac{\Gamma^2(1/4)}{4\sqrt{2\pi}}\,+\, 
{\rm Im} \left({\rm EllipticF}[{\rm ArcSin}(r/r_0), -1]\right)
\right)\,.
\label{eq:profile1}
\]
This classical trajectory gives the thin-wall bubble classical profile for $r_0 < r(\tau) <\infty$ which the result 
\eqref{eq:pdrNfinalr0} for the quantum contribution to the 
rate $\Delta W^{\rm quant}$. This trajectory is plotted in Fig.~\ref{fig:profile}.

What happens when the radius of the bubble $r(\tau)$ approaches the critical radius $r_0$ 
\eqref{r0def} where the momentum \eqref{eq:conjp} vanishes?
Recall that in the language of a mechanical analogy we are searching for an equilibrium (i.e. the stationary point 
solution) where the surface $\tau_0(r)$ located at $\tau_\infty$ at large values of $r$ is pulled upwards (in the direction of $\tau$)
by a constant force $E=nm$ acting at the point $r=0$. This is what corresponds to finding an extremum -- in our case the true minimum -- 
of the expression in \eqref{eq:halfEWqnew}, which we rewrite now in the form,
\[
\frac{1}{2} \Delta W^{\rm quant}\,=\, 
\underbrace{E\,|\tau_\infty|}_{{\rm Force}\times{\rm height}} -\,\,  \underbrace{\mu \int d^{2+1} {\rm Area}}_{\rm surface\, Energy}
\,.\label{eq:halfEWqNew}
\]
Sufficiently far away from the point at the origin where the force acts, the surface is nearly flat and does not extend in the $\tau$ direction.
As the distance in the $r$-direction closer to the point where the force is applied, the surface is getting more and more stretched in the $\tau$
direction, until the critical radius $r_0$ is reached where the the surface approaches the shape of a cylinder $R^1 \times S^2$ with $R^1$
along the $\tau$ direction. 

\begin{figure*}[t]
\begin{center}
\includegraphics[width=0.7\textwidth]{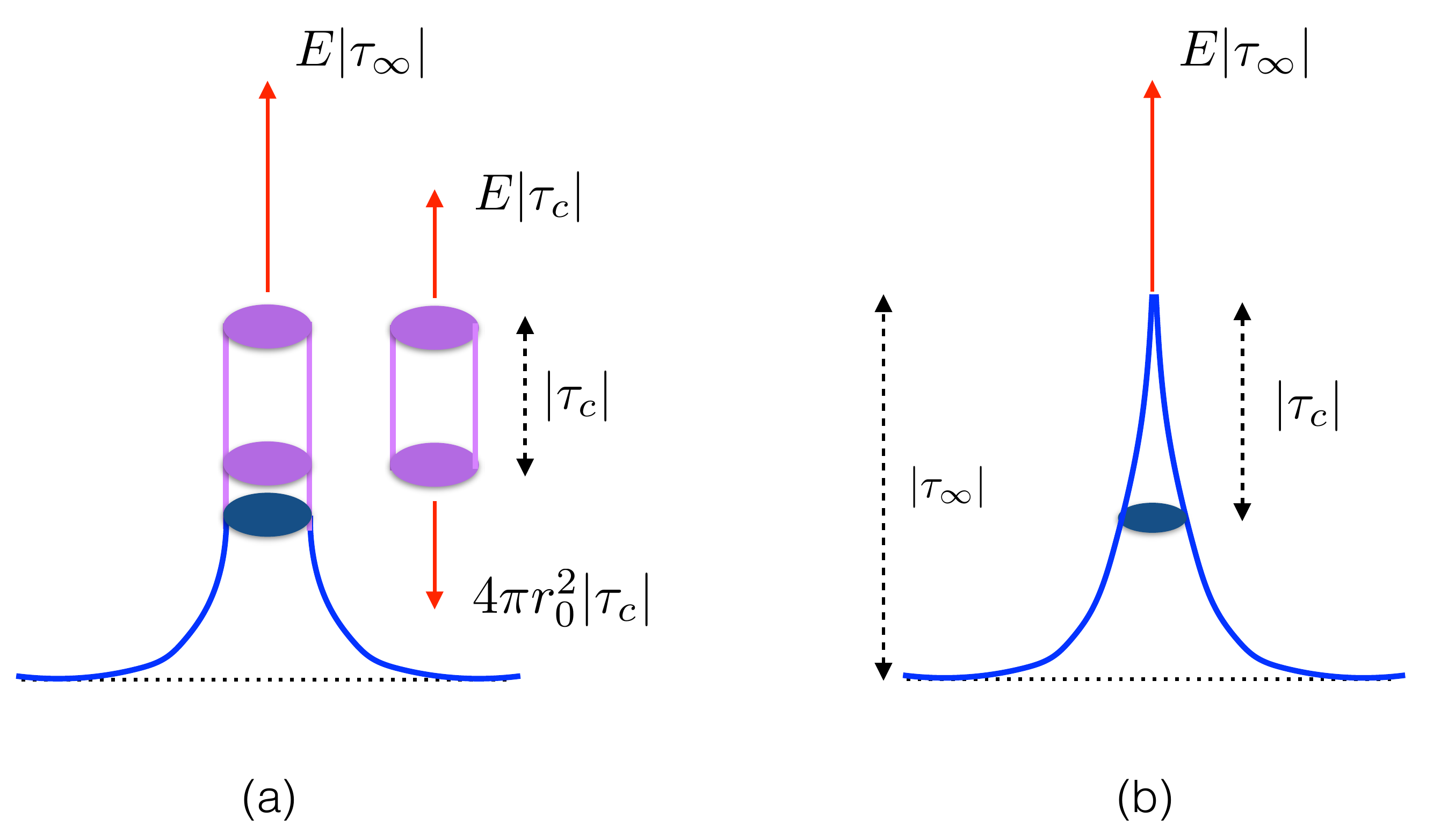}
\end{center}
\vskip-.5cm
\caption{Stationary surface configuration obtained by gluing two branches. {\bf Plot (a)}  shows the surface in the thin-wall
approximation which glues the original solution \eqref{eq:profile1} to the infinitely stretchable cylinder solution of \eqref{eq:pdrNfinalr01}.
{\bf Plot (b)} depicts its more realistic implementation where the infinite cylinder is replaced by a cone as a consequence of allowing 
the surface tension $\mu$ to increase with $|\tau|$ in the regime where the highly stretched surface becomes effectively a 1-dimensional spring. }
\label{fig:prof2}
\end{figure*}

Up to the critical point $\tau_c$ where $r=r_0$, the force and the surface tension have to balance each other in the expression,
\[
E\,|\tau_\infty-\tau_c| \,-\, \left(\int_{\tau_\infty}^{\tau_c} d\tau \,4\pi \mu \, r^2 \sqrt{1+\dot r^2}
\,-\, \frac{4\pi}{3} \,\mu R^3\right)\,,
\label{eq:pdrNfinalr01}
\]
and this is what we have calculated in Eqs.~\eqref{eq:pdrN} and \eqref{eq:pdrNfinalr0}.
But when the critical point $r_0$ is reached at a certain $\tau_c$ 
the balance of forces becomes trivial,
\[
E\,|\tau_c| \,-\, \mu \, 4\pi \,r_0^2  \, |\tau_c| \,=\, 0\,.
\label{eq:vanishing}
\]
Clearly, the branch of the classical trajectory shown as the dashed line in Fig.~\ref{fig:profile} is unphysical in the sense that it does not 
describe the membrane pulled upwards with the force $E=mn$.
The vanishing of the expression \eqref{eq:vanishing} is the consequence of the definition of the critical radius in \eqref{r0def}. 
As soon as the radius $r(\tau)$ approaches the critical radius $r_0$,
the radius freezes at this value (since $p\propto d_\tau r = 0$), 
the two terms in \eqref{eq:vanishing} become equal, $E=\,\mu \, 4\pi \,r_0^2$, and remain so at all times above the critical time $\tau_c$.
The thin-wall profile becomes an infinitely stretchable
cylinder,
as shown in Fig.~\ref{fig:prof2}~(a), giving no additional contribution to $ \Delta W^{\rm quant}$ on top of \eqref{eq:pdrNfinalr01}.

The stationary solution in the form where it becomes at $r\to r_0$ a cylinder that can be freely stretched in the vertical (i.e. $\tau$) direction
is an idealised approximation to the more realistic configuration that would be realised in our mechanical analogy of the surface stretched by the
force in practice. It is easy to see how this realistic mechanical solution looks like. For the coordinate along the vertical axis,\footnote{Recall
that the tip of the surface is at $\tau=0$ where $d=|\tau_\infty|$, and that the surface's  base is at a negative $\tau=\tau_\infty=-|\tau_\infty|$ which corresponds to $d=0$.} 
$d:= \tau +\tau_\infty \simeq 0$,
the bubble profile is nearly flat in the $\tau$ direction. As $d$ increases from $0$, the radius $r(\tau)$ grows smaller, following the profile of the thin-wall
solution contour in the lower part of Fig.~\ref{fig:prof2}. As $r$ approaches the critical radius $r_0$, the surface becomes almost entirely along the $d$ (or $\tau$)
direction. Such a surface looks more like a spring along the $\tau$ coordinate. For the strict thin-wall approximation, the surface tension $\mu$ is assumed to be
a constant. But in the case of the spring, it should be the Young's elastic modulus $k_{\rm \, Young}$ that takes a constant value. Hence for a highly stretched surface in the
$\tau$ direction we should introduce some dependence on $d=\tau+\tau_\infty$ into the surface tension,
\[
\mu \,=\, \mu_0 \,(1\,+\, \hat{k}\, (\tau+\tau_\infty)) \,,
\label{eq:newmu}
\]
where $\mu_0= \, \frac{m^3}{3\lambda}$ is the same constant contribution to the surface tension as before in \eqref{eq:mu},
and $\hat{k}\ll 1$ is a dimensionless constant. The corresponding Young's modulus of the spring-shaped stretched surface would be
$k_{\rm \,Young}= \mu_0\,\hat{k}$.
The equation \eqref{eq:newmu} describes a small deviation from the standard thin wall approximation where the surface tension
is now dependent on the stretching of the surface. This expression can be thought of as the zeroth and the first order terms in the Taylor expansion
of the function $\mu(\tau +\tau_\infty)$.
The result of this improvement on $\mu$ is that the balance between the two terms in \eqref{eq:vanishing} continues to hold, but now in the form,
\[
\left(E\,-\, \mu(d) \cdot 4\pi \,r(d)^2\right) d\,=\, 0\,, \quad {\rm where} \,\, d\ge |\tau_c| \,.
\label{eq:vanishing2}
\]
For every infinitesimal increase in the vertical coordinate $d$ above $|\tau_c|$, the radius $r(d)$ gets a little smaller than its value $r_0$ at the base of the cylinder
in Fig.~\ref{fig:prof2}~(a). As a result the cylinder gets narrower as $d$ increases and turns into the cone-like shape shown in Fig.~\ref{fig:prof2}~(b). 
The actual choice of the modification of the surface tension expression, such as in \eqref{eq:newmu}, is of course determined by the field configurations
themselves, so it can be seen as a part of the extremization procedure. One can always find an adiabatically slowly varying $\mu$ such that the contribution from the cone
to $W$ is negligible, and the overall contribution is dominated by the surface at $r>r_0$ in the large $\lambda n$ limit.
Hence we conclude that
\[
 \Delta W^{\rm quant} \,=\, 
\, \frac{1}{\lambda} \, (\lambda n)^{3/2}\, \frac{2}{\sqrt{3}}\,
\frac{\Gamma(5/4)}{\Gamma(3/4)}\,\simeq\, \, 0.854\,  n \sqrt{\lambda n}
\,.\label{eq:pdrfinal}
\]

 \bigskip
\section{Quantum rate in (2+1) dimensions}
\label{sec:lowd}
\medskip

All our calculations can be straightforwardly generalised to any number of dimensions $(d+1)$ in the same as before
scalar QFT model \eqref{eq:L} with the VEV $v\neq0$. 

The expression $W(E,N)_d$ in the exponent of the multiparticle rate ${\cal R}_n(E)$ has the same general decomposition
into the tree-level and the quantum parts as before,
\[
W(E,n)_d \,=\, W(E,n; \lambda)^{\rm tree}_d  \,+\, \Delta W(E,n; \lambda)^{\rm quant}_d\,,
\label{eq:EalWfin2d}
\]
where the tree-level expression in $(d+1)$ dimensions reads ({\it cf.} \eqref{eq:EalWfinfin}),
\[
W(E,n)^{\rm tree}_d \,=\, n \left(\log\frac{\lambda n}{4}-1\right)\,+\, \frac{dn}{2}\left( \log\frac{\varepsilon}{d\pi}+1\right)
\,,
\label{eq:EalWfinfind}
\]
and the quantum contribution is given by
\[
\Delta W^{\rm quant}_d \,=\ 2nm\,|\tau_\infty|\,+\, 2 \int  d^d x \bigg[ \int^{+\infty}_{\tau_0({\bf x})} d\tau \, {\cal L}_{\rm Eucl}(h_1) \,-\,
\int_{\tau_0({\bf x})}^0d\tau \, {\cal L}_{\rm Eucl}(h_2) \bigg]
\label{eq:EWqd233}
\]
being extremized over the singularity surfaces $\tau_0({\bf x})$
in a complete analogy with \eqref{eq:EWq}.

\medskip

For the rest of this section we we will consider the case of $d=2$ spacial dimensions and will concentrate on the contribution
of the stationary surface to the quantity $\frac{1}{2} \Delta W^{\rm quant}_{d=2}$ which we write as,
\[
\frac{1}{2}\,\Delta W^{\rm quant}_d \,=\ E\,|\tau_\infty|\,-\, 2\pi\mu \left(
\int_{r_0}^R r \sqrt{1+\dot{r}}\, dr - \int_0^R r\, dr\right)\,,
\label{eq:EWqd}
\]
where the surface tension is the same as before, $\mu =m^3/\lambda$, and the critical radius in $d=2$ is given by
$r_0 = E/(2\pi m)$. 
Proceeding with the evaluation of \eqref{eq:EWqd} on the classical trajectory $r(\tau)$ analogously to the calculation in the previous section we get,
\[
\frac{1}{2} \Delta W^{\rm quant} \,=\, 
-\, \int^{R}_{r_0}  
2\pi \, \mu \, \sqrt{r^2-r_0^2} \,dr
\,+\, {2\pi} \,\mu R^2
\,,\label{eq:pdrNd}
\]
which in the $Rm \to \infty$ limit becomes,
\[
\simeq\, \frac{n^2 \lambda}{m}\,\frac{3}{4\pi} \left(\log (Rm)\,+\, \frac{1}{2} \,+\, \log\left(\frac{2\pi}{3}\frac{m}{\lambda n}\right)
\,+\, {\cal O}\left(\frac{1}{Rm}\frac{\lambda n}{m}\right)\right)\,.
\]
Adopting the infinite volume limit where limit $Rm\to \infty$ is taken first, while the quantity $\frac{n\lambda}{m}$ is held fixed,  
we can drop the $R$-independent and $1/R$-suppressed terms, leaving only the logarithmically divergent contribution,
\[
\frac{1}{2} \Delta W^{\rm quant} \,\simeq\,  \frac{3}{4\pi}\,\frac{n^2 \lambda}{m} \log (Rm)
\label{eq:semicld2}
\]
We see that all power-like divergent terms in $mR$ have cancelled in the expressions \eqref{eq:pdrNd} and \eqref{eq:semicld2}, but the 
logarithmic divergence $\log (Rm)$ remains. This result is not surprising in $d<3$ dimensions and is the consequence of the infrared divergencies 
in the amplitudes at thresholds due to the rescattering effects of final particles. In fact, the appropriate coupling constant in the lower-dimensional theory
is not the bare coupling $\lambda$ but the running quantity $\lambda t$ where $t$ is the logarithm of the characteristic momentum scale in the final state.
In our case we can set,
\[
t\,=\, \log(Rm)
\]
and treat $R$ as one over the average momentum scale in the final state, i.e. $Rm = 1/\varepsilon^{1/2}$.

The semiclassical result obtained in \eqref{eq:semicld2} is the effect of taking into account quantum corrections to the
scattering amplitudes into $n$-particle states near their threshold, and implies
\[
A_n \,\simeq\, A_n^{\rm tree} \exp \left(\frac{3 n^2 \lambda\, t}{4\pi m} \right)\,.
\label{eq:wkb3d}
\]
It is important to recall the semiclassical limit assumed in the derivation of the above expression. It is as always
the weak-coupling plus large multiplicity limit, such that\footnote{Recall that in $(2+1)$ dimensions, $\lambda$ has dimensions of mass.}
\[
{\rm dimensionless\, running\, coupling}:\,\, \frac{\lambda\, t}{m} \to 0 \, \qquad{\rm and\,\, multiplicity}:\, n\to \infty
\]
with the quantity $ n\frac{ \lambda \,t}{m} $ held fixed (and ultimately large), and $t=-1/2\log \varepsilon \to 0$ to ensure the non-relativistic
limit which selects the amplitudes close to their multiparticle thresholds.

It is important that it is the {\rm running} coupling $\lambda t$ that is required to be small in the semiclassical exponent\footnote{For example it is completely analogous to the instanton action $S_{\rm inst} = \frac{8\pi^2}{g^2(t)}$ in the Yang-Mills theory, where the inclusion of
quantum corrections from the determinants into the instanton measure in the path integral ensures that $S_{\rm inst}$ in the exponent depends on the correct RG coupling 
$g^2(t)$ and not the unphysical bare coupling $g^2_{\rm bare}$.}.
This implies that the semiclassical expression would in general include unknown corrections in
 \[
A_n \,\simeq\, A_n^{\rm tree} \exp \left(\frac{3 n^2 \lambda\, t}{4\pi m}\left(1+ \sum_{k=1}^\infty c_k\left(\frac{\lambda t}{m} \right)^k\right) \right)\,,
\label{eq:wkb3dsum}
\] 
parameterised by the sum $\sum_{k=1} c_k\left(\frac{\lambda t}{m} \right)^k$. Of course, there is a well-defined regime corresponding to 
the small values of the effective coupling $\lambda t$ where these
corrections are negligible and the leading order semiclassical result in \eqref{eq:wkb3d} is justified.

\medskip

Remarkably, the semiclassical formula \eqref{eq:wkb3d} can be tested against an independent computation of quantum effects
in the $(2+1)$-dimensional theory obtained in~\cite{Rubakov:1994cz,Libanov:1994ug} using the RG resummation of perturbative diagrams.
The result is,
\[
A_n^{\rm RG} \,=\, A_n^{\rm tree} \,\left(1\,-\, \frac{3 \lambda\, t}{2\pi m} \right)^{-\, \frac{n(n-1)}{2}}\,.
\label{eq:RG2d}
\]
This expression is supposed to be valid for any values of $n$, and in the regime where the effective coupling $\lambda t$
is in the interval,
\[
0\, \le \, \frac{\lambda\, t}{m}\, \lesssim\, 1\,.
\]
Now taking the large-$n$ limit the RG-technique based result of~\cite{Rubakov:1994cz,Libanov:1994ug} gives
\[
A_n^{\rm RG} \,=\, A_n^{\rm tree} \,\exp \left(\frac{3 n^2 \lambda\, t}{4\pi m}\left(1+ \sum_{k=1}^\infty \frac{1}{k+1}\,\left(\frac{3\lambda t}{2\pi m} \right)^k\right) \right)\,,
\label{eq:RG2dT}
\]
It is a nice test of the semiclassical approach that the leading order terms in the exponent in both expressions, \eqref{eq:wkb3dsum} and \eqref{eq:RG2dT}
are exactly the same and given by $\frac{3 n^2 \lambda\, t}{4\pi m}$.
An equally important observation is that the subleading terms are of the form $\sum_{k=1} c_k\left(\frac{\lambda t}{m} \right)^k$ which is suppressed in the 
semiclassical limit $\lambda t\to 0$. There is no contradiction between the two expressions in the regime where the semiclassical approach is justified.

\medskip

It thus follows that there is a regime in the $(2+1)$-dimensional theory where the multiparticle amplitudes near their thresholds, and consequently 
the probabilistic rates ${\cal R}_n(E)$ become large.
In the case of the RG expression \eqref{eq:RG2d}, this is the consequence of taking a large negative power $-n^2/2$ of the term that is smaller than 1.
This implies that there is a room for realising Higgsplosion in this $(2+1)$-dimensional model in the broken phase. 

In the case of a much simpler model -- the quantum mechanical anharmonic oscillator in the unbroken phase -- it was recently shown in Ref.~\cite{Jaeckel:2018ipq}
that the rates remain exponentially suppressed in accordance with what would be expected from unitarity in QM.

 \bigskip
\section{Conclusions}
\label{sec:concl}
\bigskip

In this paper, following the idea outlined in our earlier work \cite{Khoze:2017ifq}
we computed the semiclassical exponent of the multi-particle 
production rate in the high-particle-number $\lambda n \to \infty$ 
limit in the kinematical regime where the final state particles are produced near their mass thresholds.
This corresponds to the limit 
\[ \lambda \to 0\,, \quad n\to \infty\,, \quad {\rm with}\quad
\lambda n = {\rm fixed} \gg 1 \,, \quad \varepsilon ={\rm fixed} \ll 1 \,.
\label{eq:limit2}
\]
Combining the tree-level \eqref{eq:EalWfinfin} and the quantum effects \eqref{eq:pdrfinal}
contributions,
\[
W(E,n) \,=\, W(E,n; \lambda)^{\rm tree}  \,+\, \Delta W(E,n; \lambda)^{\rm quant}\,,
\label{eq:EalWfin2222}
\]
we can write down the full semiclassical rate,
\[
{\cal R}_n(E)\,= \, e^{W(E,n)}\,=\, 
\exp \left[ n\, \left( 
\log \frac{\lambda n}{4} \,+\, 0.85\, \sqrt{\lambda n}\,+\,\frac{3}{2}\log \frac{\varepsilon}{3\pi} \,+\, \frac{1}{2}
\right)\right] 
\label{eq:Rnp2}
\]
computed in the high-multiplicity non-relativistic limit \eqref{eq:limit2}.
This expression for the multi-particle rates was first written down in the precursor of this work \cite{Khoze:2017ifq}, and
was used in Refs.~\cite{Khoze:2017tjt,Khoze:2017lft} and subsequent papers to introduce and motivate the Higgsplosion mechanism.

\begin{figure*}[t]
\begin{center}
\includegraphics[width=1.0\textwidth]{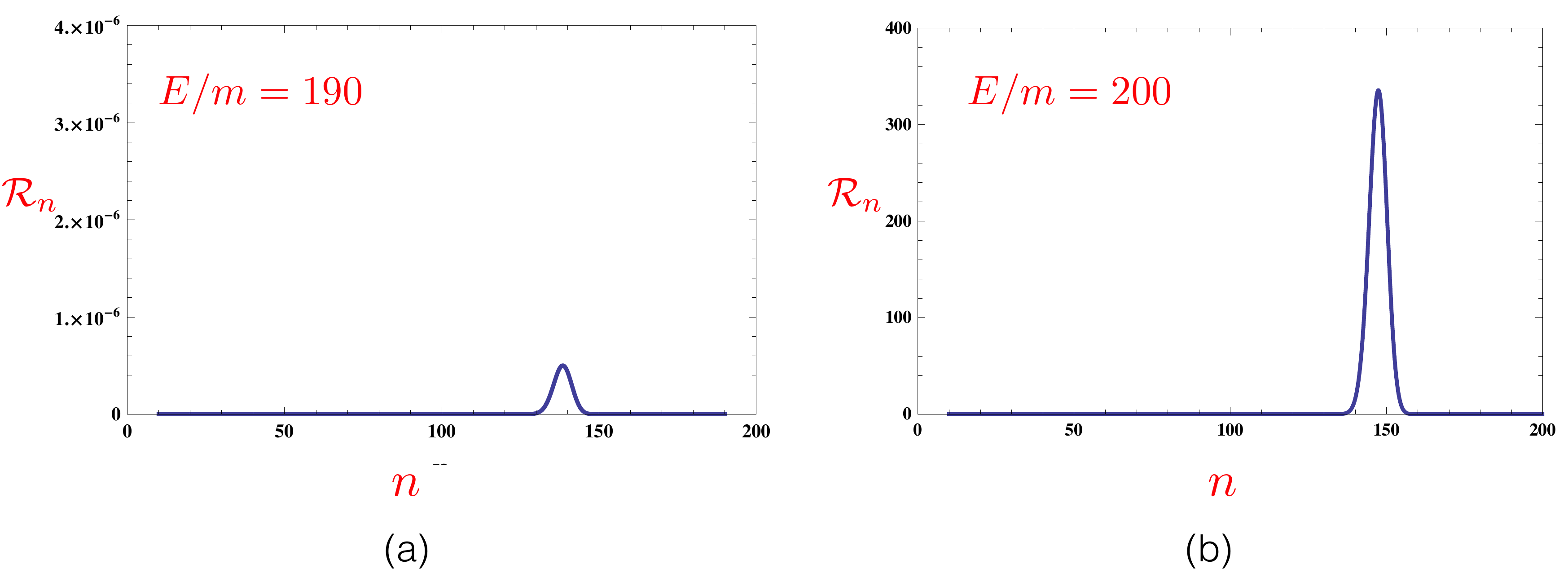}
\end{center}
\vskip-.5cm
\caption{Plots of the semiclassical rate ${\cal R}_n$ in Eq.~\eqref{eq:Rnp2}
 as a function of $n$ for values of the energy/virtuality $E$ fixed at 
190$m$ and at 200$m$. We chose $\lambda=1/8$. There is a sharp exponential dependence of the peak rate
on the energy. The peak multiplicities $n \sim 150$ in these examples are not far below the maximal values $n_{\rm max} = E/m$
allowed by kinematics.
 } 
\label{fig:R2}
\end{figure*}

The energy in the initial state and the final state multiplicity are related linearly via
\[ E/m \,=\, (1 + \varepsilon) \, n\,,
\]
and thus for any fixed non-vanishing value of $\varepsilon$, one can raise the energy to 
achieve any desired large value of $n$ and consequentially a large $\sqrt{\lambda n}$.
Clearly, at the strictly vanishing value of $\varepsilon$, the phase-space volume is zero and 
the entire rate \eqref{eq:Rnp2} vanishes. Then by increasing $\varepsilon$ to a positive but still small values, the rate increases.
The competition is between the negative $\log \varepsilon$ term and the positive $\sqrt{\lambda n}$ term 
in \eqref{eq:Rnp2}, and there is always a range of sufficiently high multiplicities where $\sqrt{\lambda n}$
overtakes the logarithmic term $\log \varepsilon$ for any fixed (however small) value of $\varepsilon$.
This leads to the exponentially growing multi-particle rates above a certain critical energy, which in the case
described by the expression in \eqref{eq:Rnp2} is in the regime of $E_c \sim 200 m$.
We refer the reader to Fig.~\ref{fig:R2} and to section 5 of Ref.~\cite{Khoze:2017ifq} for a detailed discussion of the exponential rate 
\eqref{eq:Rnp2} and its relevance for Higgsplosion~\cite{Khoze:2017tjt}.

\bigskip

Our discussion concentrated entirely on a simple scalar QFT model.
If more degrees of freedom were included, for example the $W$ and $Z$ vector bosons and
the SM fermions, new coupling parameters (such as the gauge coupling and the Yukawas) 
would appear in the expression for the rate along with the final state particle multiplicities. As there are more parameters,
the simple scaling properties of ${\cal R}_n$ in the pure scalar theory will be modified. Understanding how this would work in practice and 
investigating the appropriate semiclassical limits is one (of the admittedly many) tasks for 
future work on exploring realisations of Higgsplosion in particle physics. 

\bigskip
\bigskip

\section*{Acknowledgements}

I am grateful to Joerg Jaeckel, Joey Reiness, Jakub Scholtz and Michael Spannowsky for many useful discussions.

\bigskip

\appendix

\section{Appendix: Comments on the semiclassical method}
\label{app:Son}
\medskip

The aim of the semiclassical approach of~\cite{Son:1995wz} is to compute the probability rate ${\cal R}_n(E)$ for a local operator ${\cal O}(x)$ at a point $x=0$
to create $n$ particles of total energy $E$ from the vacuum,
\[
{\cal R}_n(E)\,=\, \int d\Phi_n\, \langle 0| \,{\cal O}^\dagger \, S^\dagger\, P_E |n\rangle 
\langle n|\, P_E\, S\, {\cal O}\, |0\rangle\,.
\label{Aeq:RnE1}
\]
The matrix element is squared and integrated over the 
$n$-particle Lorentz-invariant phase space $\Phi_n$
\[ 
\int d\Phi_{n} \,=\, \frac{1}{n!}\,(2\pi)^4 \delta^{(4)}(P_{\rm in}-\sum_{j=1}^n p_j) \,
 \prod_{j=1}^n \int \frac{d^3 p_j}{(2\pi)^3\, 2 p_j^0} \,.
 \label{Aeq:phase-sp}
\]
Note that in our conventions 
the bosonic phase-space volume element \eqref{Aeq:phase-sp} 
includes the $1/n!$ symmetry factor for the production of the $n$ equivalent Higgs bosons.\footnote{Hence the $n$-particle 
cross-sections ${\cal R}_n(E)$ still retains a single factor of $n!$. Indeed, according to \eqref{eq:ampln2},
the amplitude squared contributes the factor of $(n!)^2$, and combining with the symmetry factor from the 
bosonic $n$-particle phase space 
we have ${\cal R}_n(E) \sim\, \frac{1}{n!} \, n! \, n! \sim \, n!$.
}

The original Landau WKB method \cite{Landau} was setup for computing
matrix elements of generic operators in Quantium Mechanics between the initial and final states with different 
energy eigenvalues. 
In the QFT settings,
the initial state is a vacuum and the final state is the $n$-particle final state with $n\gg 1$.
It is known that to the leading exponential accuracy the transition rates computed using the Landau WKB method do not
depend on the specific form of the operator ${\cal O}$ used to deform the initial state, if this deformation is not exponential. 
It is then similarly expected that 
the choice of the operator in \eqref{eq:opdef} does not affect the exponent in the transition rates in the QFT settings either.

The multiparticle rate \eqref{Aeq:RnE1} in question is represented as the double functional integral 
(one for each of the matrix elements) with additional integrations over the Lagrange multipliers implementing the 
projections onto final states with finite energy and particle number. All these integrals
are subsequently computed in the steepest descent approximation
for all integration variables which is justified in the 
double-scaling weak-coupling / large-$n$ semiclassical limit \eqref{eq:limit}.

The semiclassical result for the rate then takes the form \cite{Son:1995wz} (for an extensive recent review of the semiclassical
method and the derivation of the results quoted below see Ref.~\cite{Khoze:2018mey})
\begin{eqnarray}
{\cal R}_n(E) &\simeq& \exp \left[ W(E,n)\right]\,
\label{Aeq:ReW}
\\
W(E,n) &\equiv& \frac{1}{\lambda}\, {\cal F}(\lambda n, \varepsilon)\,=\,\,
ET \,-\, n\theta \,-\, 2{\rm Im} S[h]\,.
\label{Aeq:Wdef}
\end{eqnarray}
Let us now examine the structure of this result.
The function ${\cal F}(\lambda n, \varepsilon)$ appearing in \eqref{Aeq:Wdef}, is a function of two arguments, $\lambda n$ and $\varepsilon$, 
characterising the final $n$-particle state with the average kinetic energy per particle per mass $\varepsilon$.
All the integrations in the path integral representation of ${\cal R}_n(E)$ in \eqref{Aeq:RnE1}  were carried out 
and saturated by their saddle-point values in the large-$n$, large-$1/\lambda$ limit \eqref{eq:limit}.
At negative values of ${\cal F}(\lambda n, \varepsilon)$ the multi-particle rate ${\cal R}_n(E)$ is exponentially suppressed,
while if ${\cal F}(\lambda n, \varepsilon)$ crosses zero and becomes positive above some critical energy or multiplicity, the
multi-particle processes enter the Higgsplosion phase \cite{Khoze:2017tjt}.

\bigskip

\noindent The function $W(E,n)$  is computed on the saddle-point value of the path integral.
We now consider the terms appearing in the final expression in \eqref{Aeq:Wdef}.
First, the combination $-\, 2{\rm Im} S[h]$ follows from the $e^{-iS^*} e^{iS}$ factor (where $S$ is the action)
in the product of the matrix elements in  \eqref{eq:RnE1}. The integration contours and the 
resulting saddle-points in the steepest descent integration are complex-valued, 
hence $iS[h]-iS[h]^*\, =\, -\, 2{\rm Im} S[h]$
or equivalently $-2S_{\rm Eucl}[h]$ using the Euclidean notation.
The remaining parameters, $T$ and $\theta$, appearing on the right hand side of \eqref{Aeq:Wdef},  are the Lagrange multipliers
that emerged from the projection operators $P_E$ and $P_n$ onto the final states with defined values of the energy $E$ 
and the particle number $n$ 
in \eqref{Aeq:RnE1}. The parameters $T$ and $\theta$ are some of the integration variables in the integral representation of \eqref{Aeq:RnE1};
in the steepest descent approximation, they form a part of the saddle point parameter set and take the fixed value on a given saddle point solution.

Prior to taking the $j\to 0$ limit, the
saddle-point field configuration $h(x)$ is given by a particular solution to the classical equation
of motion with the singular 
source term $j(x)=\, j \delta^{(4)}(x)$ on the right hand side,
\[
\frac{\delta S}{\delta h(x)}\,=\, i\, j\, \delta^{(4)}(x)\,,
\label{sing_eqn}
\]
where $S=\int d^4 x {\cal L}$ is the action of the theory and $j$ is a constant.
After taking the limit $j\to 0$, the right hand side of the defining equation \eqref{sing_eqn} vanishes
but the required solution nevertheless remains singular at $x=0$ in Minkowski space.
The saddle-point solution also depends on the parameters $T$ and $\theta$, as will be explained below, while the 
overall expression $W(E,n)$ is independent of  $T$ and $\theta$. Hence,
\[ 
2\, \frac{\partial\,{\rm Im}S}{\partial T}\,=\, E\,, \qquad
2\, \frac{\partial\,{\rm Im}S}{\partial \theta}\,=\, -\,n\,,
\label{eqs:Tthetadef}
\]
and $W(E,n)$ is the Legendre transformation of the action $2{\rm Im}S$ 
with respect to $T$ and $\theta$.\footnote{Indeed, it follows from the definition of $W$ that
$\frac{\partial W}{\partial E}\,=\, T$ and 
$\frac{\partial W}{\partial n}\,=\, -\,\theta$.
The action $S[h]$ depends on the parameters $T$ and $\theta$ through the classical solution $h(x)$,
but in the final expression for $W(E,n)$ these parameters are traded for $E$ and $n$. }

Next step is to specify the boundary conditions of the solution $h(x)$ at  $t_{\rm in}\to -\infty$ and $t_{\rm fin}\to +\infty$.
At the initial and final time boundaries $h(x)$ satisfies the free Klein-Gordon equation, thus
\begin{eqnarray}
h({\bf x},t)|_{t\to - \infty} &\to& v\,+\, 
\int \frac{d^3k}{(2\pi)^{3/2}} \frac{1}{\sqrt{2\omega_{\bf k}}}\,\, a^*_{\bf k}\, e^{ik_\mu x^\mu} \,
\label{eq:limin}
\\
h({\bf x},t)|_{t\to + \infty} &\to& v\,+\, 
\int \frac{d^3k}{(2\pi)^{3/2}} \frac{1}{\sqrt{2\omega_{\bf k}}}\left(c_{\bf k}\, e^{-ik_\mu x^\mu}\,+\, b^*_{\bf k}\, e^{ik_\mu x^\mu}\right)\,.
\label{eq:limf}
\end{eqnarray}
where we used the standard notation $k_0=\omega_{\bf k}=\sqrt{m^2+{\bf k}^2}$
so that $e^{\pm i k_\mu x^\mu} = e^{\pm i(\omega_{\bf k} t \,-\, {\bf k} {\bf x})}$. \\
The $t \to -\infty$ boundary condition in Eq.~\eqref{eq:limin} 
contains only the positive frequency components
$ a^*_{\bf k} \, e^{-i\omega_{\bf k} |t|}$ and no negative frequency ones 
$ a_{\bf k} \, e^{+i\omega_{\bf k} |t|}$.
In the second quantisation operator formalism, this condition
implements the requirement that there are no particles in the initial state, 
since the creation operator $\hat{a}^\dagger$ annihilates the bra-state vacuum $\langle 0|.$
The second boundary condition \eqref{eq:limf} at the final time $t\to +\infty$
contains both positive and negative frequency components.
Following \cite{Son:1995wz} we parameterise its $c_{\bf k}$ coefficient  
in terms of the complex conjugate of its $b_{\bf k}^\dagger$ coefficient,
\[
c_{\bf k}\,=\, b_{\bf k}\,e^{\omega_{\bf k}T-\theta}\,.
\label{eq:cb}
\]
The solution is complex-valued since $c_{\bf k}\neq\, b_{\bf k}$,
and the corresponding parameters $T$ and $\theta$ are precisely those appearing in \eqref{eqs:Tthetadef}. 
For more detail we refer the reader to \cite{Khoze:2018mey}.

\bigskip

\bibliographystyle{JHEP}
\bibliography{references}

\providecommand{\href}[2]{#2}\begingroup\raggedright\begin{thebibliography}{10}

\bibitem{Son:1995wz}
D.~T. Son, \emph{{Semiclassical approach for multiparticle production in scalar
  theories}}, \href{http://dx.doi.org/10.1016/0550-3213(96)00386-0}{\emph{Nucl.
  Phys.} {\bf B477} (1996) 378--406},
  [\href{https://arxiv.org/abs/hep-ph/9505338}{{\tt hep-ph/9505338}}].

\bibitem{Khoze:2017ifq}
V.~V. Khoze, \emph{{Multiparticle production in the large $\lambda n$ limit:
  realising Higgsplosion in a scalar QFT}},
  \href{http://dx.doi.org/10.1007/JHEP06(2017)148}{\emph{JHEP} {\bf 06} (2017)
  148}, [\href{https://arxiv.org/abs/1705.04365}{{\tt 1705.04365}}].

\bibitem{Khoze:2017tjt}
V.~V. Khoze and M.~Spannowsky, \emph{{Higgsplosion: Solving the Hierarchy
  Problem via rapid decays of heavy states into multiple Higgs bosons}},
  \href{http://dx.doi.org/10.1016/j.nuclphysb.2017.11.002}{\emph{Nucl. Phys.}
  {\bf B926} (2018) 95--111}, [\href{https://arxiv.org/abs/1704.03447}{{\tt
  1704.03447}}].

\bibitem{Khoze:2017lft}
V.~V. Khoze and M.~Spannowsky, \emph{{Higgsploding universe}},
  \href{http://dx.doi.org/10.1103/PhysRevD.96.075042}{\emph{Phys. Rev.} {\bf
  D96} (2017) 075042}, [\href{https://arxiv.org/abs/1707.01531}{{\tt
  1707.01531}}].

\bibitem{Gainer:2017jkp}
J.~S. Gainer, \emph{{Measuring the Higgsplosion Yield: Counting Large Higgs
  Multiplicities at Colliders}},  \href{https://arxiv.org/abs/1705.00737}{{\tt
  1705.00737}}.

\bibitem{Khoze:2017uga}
V.~V. Khoze, J.~Reiness, M.~Spannowsky and P.~Waite, \emph{{Precision
  measurements for the Higgsploding Standard Model}},
  \href{https://arxiv.org/abs/1709.08655}{{\tt 1709.08655}}.

\bibitem{Khoze:2018bwa}
V.~V. Khoze, J.~Reiness, J.~Scholtz and M.~Spannowsky, \emph{{A Higgsploding
  Theory of Dark Matter}},  \href{https://arxiv.org/abs/1803.05441}{{\tt
  1803.05441}}.

\bibitem{Gorsky:1993ix}
A.~S. Gorsky and M.~B. Voloshin, \emph{{Nonperturbative production of
  multiboson states and quantum bubbles}},
  \href{http://dx.doi.org/10.1103/PhysRevD.48.3843}{\emph{Phys. Rev.} {\bf D48}
  (1993) 3843--3851}, [\href{https://arxiv.org/abs/hep-ph/9305219}{{\tt
  hep-ph/9305219}}].

\bibitem{Cornwall:1990hh}
J.~M. Cornwall, \emph{{On the High-energy Behavior of Weakly Coupled Gauge
  Theories}}, \href{http://dx.doi.org/10.1016/0370-2693(90)90850-6}{\emph{Phys.
  Lett.} {\bf B243} (1990) 271--278}.

\bibitem{Goldberg:1990qk}
H.~Goldberg, \emph{{Breakdown of perturbation theory at tree level in theories
  with scalars}},
  \href{http://dx.doi.org/10.1016/0370-2693(90)90628-J}{\emph{Phys. Lett.} {\bf
  B246} (1990) 445--450}.

\bibitem{Brown:1992ay}
L.~S. Brown, \emph{{Summing tree graphs at threshold}},
  \href{http://dx.doi.org/10.1103/PhysRevD.46.R4125}{\emph{Phys. Rev.} {\bf
  D46} (1992) R4125--R4127}, [\href{https://arxiv.org/abs/hep-ph/9209203}{{\tt
  hep-ph/9209203}}].

\bibitem{Argyres:1992np}
E.~N. Argyres, R.~H.~P. Kleiss and C.~G. Papadopoulos, \emph{{Amplitude
  estimates for multi - Higgs production at high-energies}},
  \href{http://dx.doi.org/10.1016/0550-3213(93)90140-K}{\emph{Nucl. Phys.} {\bf
  B391} (1993) 42--56}.

\bibitem{Voloshin:1992rr}
M.~B. Voloshin, \emph{{Estimate of the onset of nonperturbative particle
  production at high-energy in a scalar theory}},
  \href{http://dx.doi.org/10.1016/0370-2693(92)90901-F}{\emph{Phys. Lett.} {\bf
  B293} (1992) 389--394}.

\bibitem{Voloshin:1992nu}
M.~B. Voloshin, \emph{{Summing one loop graphs at multiparticle threshold}},
  \href{http://dx.doi.org/10.1103/PhysRevD.47.R357}{\emph{Phys. Rev.} {\bf D47}
  (1993) R357--R361}, [\href{https://arxiv.org/abs/hep-ph/9209240}{{\tt
  hep-ph/9209240}}].

\bibitem{Libanov:1994ug}
M.~V. Libanov, V.~A. Rubakov, D.~T. Son and S.~V. Troitsky,
  \emph{{Exponentiation of multiparticle amplitudes in scalar theories}},
  \href{http://dx.doi.org/10.1103/PhysRevD.50.7553}{\emph{Phys. Rev.} {\bf D50}
  (1994) 7553--7569}, [\href{https://arxiv.org/abs/hep-ph/9407381}{{\tt
  hep-ph/9407381}}].

\bibitem{Libanov:1997nt}
M.~V. Libanov, V.~A. Rubakov and S.~V. Troitsky, \emph{{Multiparticle processes
  and semiclassical analysis in bosonic field theories}},
  \href{http://dx.doi.org/10.1134/1.953038}{\emph{Phys. Part. Nucl.} {\bf 28}
  (1997) 217--240}.

\bibitem{Khoze:2014zha}
V.~V. Khoze, \emph{{Multiparticle Higgs and Vector Boson Amplitudes at
  Threshold}}, \href{http://dx.doi.org/10.1007/JHEP07(2014)008}{\emph{JHEP}
  {\bf 07} (2014) 008}, [\href{https://arxiv.org/abs/1404.4876}{{\tt
  1404.4876}}].

\bibitem{Khoze:2014kka}
V.~V. Khoze, \emph{{Perturbative growth of high-multiplicity W, Z and Higgs
  production processes at high energies}},
  \href{http://dx.doi.org/10.1007/JHEP03(2015)038}{\emph{JHEP} {\bf 03} (2015)
  038}, [\href{https://arxiv.org/abs/1411.2925}{{\tt 1411.2925}}].

\bibitem{Jaeckel:2014lya}
J.~Jaeckel and V.~V. Khoze, \emph{{Upper limit on the scale of new physics
  phenomena from rising cross sections in high multiplicity Higgs and vector
  boson events}},
  \href{http://dx.doi.org/10.1103/PhysRevD.91.093007}{\emph{Phys. Rev.} {\bf
  D91} (2015) 093007}, [\href{https://arxiv.org/abs/1411.5633}{{\tt
  1411.5633}}].

\bibitem{Khoze:2015yba}
V.~V. Khoze, \emph{{Diagrammatic computation of multi-Higgs processes at very
  high energies: Scaling log $\sigma_n$ with MadGraph}},
  \href{http://dx.doi.org/10.1103/PhysRevD.92.014021}{\emph{Phys. Rev.} {\bf
  D92} (2015) 014021}, [\href{https://arxiv.org/abs/1504.05023}{{\tt
  1504.05023}}].

\bibitem{Degrande:2016oan}
C.~Degrande, V.~V. Khoze and O.~Mattelaer, \emph{{Multi-Higgs production in
  gluon fusion at 100 TeV}},
  \href{http://dx.doi.org/10.1103/PhysRevD.94.085031}{\emph{Phys. Rev.} {\bf
  D94} (2016) 085031}, [\href{https://arxiv.org/abs/1605.06372}{{\tt
  1605.06372}}].

\bibitem{Jaeckel:2018ipq}
J.~Jaeckel and S.~Schenk, \emph{{Exploring High Multiplicity Amplitudes in
  Quantum Mechanics}},  \href{https://arxiv.org/abs/1806.01857}{{\tt
  1806.01857}}.

\bibitem{Landau}
L.~D. Landau and E.~M. Lifshitz, \emph{{Quantum Mechanics}}, {\emph{Pergamon
  Press} (1977) }.

\bibitem{Landau2}
L.~D. Landau, \emph{{On the theory of transfer of energy at collisions I}},
  {\emph{Phys. Zs. Sowiet.} {\bf 1} (1932) 88}.

\bibitem{IP}
S.~V. Iordanskii and L.~P. Pitaevskii, \emph{{Multiphoton boundary of the
  excitation spectrum in He II}}, {\emph{Sov. Phys. JETP} {\bf 49} (1979) 386}.

\bibitem{Voloshin:1990mz}
M.~B. Voloshin, \emph{{On strong high-energy scattering in theories with weak
  coupling}}, \href{http://dx.doi.org/10.1103/PhysRevD.43.1726}{\emph{Phys.
  Rev.} {\bf D43} (1991) 1726--1734}.

\bibitem{Khlebnikov:1992af}
S.~{\relax Yu}. Khlebnikov, \emph{{Semiclassical approach to multiparticle
  production}},
  \href{http://dx.doi.org/10.1016/0370-2693(92)90669-U}{\emph{Phys. Lett.} {\bf
  B282} (1992) 459--465}.

\bibitem{Diakonov:1993ha}
D.~Diakonov and V.~Petrov, \emph{{Nonperturbative isotropic multiparticle
  production in Yang-Mills theory}},
  \href{http://dx.doi.org/10.1103/PhysRevD.50.266}{\emph{Phys. Rev.} {\bf D50}
  (1994) 266--282}, [\href{https://arxiv.org/abs/hep-ph/9307356}{{\tt
  hep-ph/9307356}}].

\bibitem{Khoze:2018mey}
V.~V. Khoze and J.~Reiness, \emph{{Review of the semiclassical formalism for
  multiparticle production at high energies}},
  \href{https://arxiv.org/abs/1810.01722}{{\tt 1810.01722}}.

\bibitem{Rubakov:1994cz}
V.~A. Rubakov and D.~T. Son, \emph{{Renormalization group for multiparticle
  production in (2+1)-dimensions around the threshold}},  in \emph{{8th
  International Seminar on High-energy Physics (Quarks 94) Vladimir, Russia,
  May 11-18, 1994}}, pp.~233--240, 1994.
\newblock \href{https://arxiv.org/abs/hep-ph/9406362}{{\tt hep-ph/9406362}}.

\end{thebibliography}\endgroup

\end{document}